\documentclass[aps,prb,twocolumn,groupedaddress, showpacs, showkeys]{revtex4}
\usepackage[]{graphicx}
\usepackage{subfigure}

\newcommand{\MAB}{Mg$_{1-x}$Al$_{x}$B$_2$}

\newcommand{\MAbB}{Mg$_{0.85}$Al$_{0.15}$B$_2$}
\newcommand{\MAcB}{Mg$_{0.75}$Al$_{0.25}$B$_2$}

\newcommand{\MABC}{Mg$_{1-x}$Al$_{x}$(B$_{1-y}$C$_{y}$)$_2$}
\newcommand{\MBCa}{Mg(B$_{0.96}$C$_{0.04}$)$_2$}
\newcommand{\MBCb}{Mg(B$_{0.92}$C$_{0.08}$)$_2$}

\newcommand{\MBC}{Mg(B$_{1-y}$C$_{y}$)$_2$}
\newcommand{\MB}{MgB$_2$}

\newcommand{\DEG}{$^{\circ}\ $}

\newcommand{\Ss}{$\sigma^\ast$}
\newcommand{\Sig}{$\sigma$}
\newcommand{\PI}{$\pi$}
\newcommand{\HT}{$H_{c_2}$}

\begin{document}

\preprint{Phys Rev B/MgB$_2$}
\title{Electron energy-loss spectroscopy study of electron-doping in MgB$_2$}

\author{R.~F.~Klie}
\email[]{klie@bnl.gov}
\author{J.~C.~Zheng}
\author{Y.~Zhu}
\affiliation{Brookhaven National Laboratory, Center for Functional Nanomaterials, Upton, NY 11973}

\author{A.~J.~Zambano}
\author{L.~D.~Cooley}

\affiliation{Brookhaven National Laboratory, Dept.\ of Materials
Science, Upton, NY 11973}

\date{\today}

\begin{abstract}
The electronic structure of electron-doped polycrystalline Mg$_{1-x}$Al$_{x}$(B$_{1-y}$C$_{y}$)$_2$ was examined by electron energy-loss spectroscopy (EELS) in a scanning transmission electron microscope (STEM) and first-principle electronic structure calculations. We found significant changes in the boron $K$ edge fine structure, suggesting the two bands of the B $K$ edge, the $\sigma$ and the $\pi$ band are being simultaneously filled as the electron doping concentration of Mg$_{1-x}$Al$_{x}$(B$_{1-y}$C$_{y}$)$_2$ was increased.
Our density-functional theory calculations confirm the filling of the $\sigma$ band states close to the Fermi level, which is believed to cause the loss of  superconductivity in highly doped MgB$_2$, since the electron-phonon coupling of these states is thought to be responsible for the high superconducting transition temperature.
Our results do not show significant differences in the electronic structure for electron doping on either the Mg or the B site, although many superconducting properties, such as $T_c$ or H$_{c_2}$ differ considerably for C- and Al- doped \MB. This behavior can not be satisfactorily explained by band filling alone, and effects such as interband scattering are  considered.

\end{abstract}


\maketitle

\section*{Introduction}
Since its  discovery as a conventional BCS superconductor with an unexpectedly high superconducting transition temperature ($T_c$) of 39 K, \cite{Nagamatsu01} \MB\ has been studied intensely over the last four years, and many of its  properties  are now well understood. As such, it was established very early that \MB\ has four separate sheets at the Fermi surface, two of them being degenerate and corresponding to a 2-dimensional \Sig\ band and two comprising 3-dimensional \PI\ bands. \cite{Brinkman02, LiuPRL01} The different electron-phonon coupling strengths of these bands lead to two distinct superconducting gaps, with the \Sig\ band gap having the higher electron-phonon coupling and thus being the main contributor to the unusually high $T_c$. It was further shown that $T_c$ could be decreased significantly by replacing Mg with Al or B with C, \cite{Agrestini01, Papagelis03, Schmidt03, DiCastro02, Postorino02, Bianconi02, Zambano05, Papavassiliou02, Margadonna02, Putti03, Putti05} which was explained by many different mechanisms ranging from band-filling,\cite{Kortus05} to a merging of the superconducting gaps,\cite{Gonnelli05} to increased interband scattering and a decrease in the electron-phonon coupling strength.\cite{Matsui04}
While many different mechanisms are now being discussed in the literature, there has not yet been any consistent explanation for the loss of $T_c$ for Al- or C doped \MB. It was previously shown that adding electrons to \MB, by doping with Al or C, reduces the bond length (i.e. the unit-cell volume decreases as a function of doping concentration), which lowers the density of states (DOS) and simultaneously stiffens the available phonon modes.\cite{Matsui04} However, the effect of decreased lattice parameters, as evidenced by measurements of $T_c$ vs pressure does not by itself explain the full $T_c$ decrease measured for Al- or C-doped \MB.\cite{Tomita01} Alternative mechanisms for the decrease of $T_c$, such as increased interband scattering due to the doping atoms or band filling, have been proposed but no experimental evidence has been reported. Indeed, initial band structure calculations for Al- and C doped \MB\ suggested that the decrease in $T_c$ can be understood by simple band-filling, caused by the extra electrons provided to the system.\cite{Kortus05} Although the effects of Al- and C doping on $T_c$ and the superconducting gaps seem to be remarkably similar, other superconducting properties, such as the upper critical field (\HT) show completely different behavior. In particular, it was shown that \HT\ is dramatically increased in both poly-crystalline \MBC\, and especially in thin-film samples.\cite{Pogrebnyakov04} On the other hand, \MAB\ shows only a weak increase in \HT\ for small concentrations of Al, and otherwise a significant decrease.\cite{Karpinski05}

Kortus et al.\cite{Kortus05} pointed out that the experimental data on both doped and ``pure'' \MB\ show a large spread of properties, e.g. $T_c$ of 10\% C-doped \MB\ ranges between 12K and 34K depending on the synthesis and annealing conditions. Moreover, the exact doping concentration of the \MB\ grains is not measured directly in most studies, but rather measurements that average over many grains are performed.
Therefore, a systematic study of the effects of C- and Al doping on the electronic structure and the superconducting properties, with doping concentrations measured inside the \MABC\ grains, is needed to fully understand its complex structure-property relationships. Also, most conventional electronic-structure measurements, such as X-ray absorption spectroscopy (XAS), cannot study individual grains or distinguish between different crystallographic orientations, since no \textbf{large} single crystals of Al- or C doped \MB\ have been available to date. Thus, we need to use electronic structure probes that are significantly smaller than the grain size of \MABC\ in order to accurately measure the effects of doping on the local DOS and its anisotropy.

In this paper, we explore the electronic structure of Al- and C doped \MB\ by combined high-resolution Z-contrast imaging and electron energy-loss spectroscopy (EELS) in a scanning transmission electron microscope (STEM). Our previous work \cite{klie03, klie02APL, zhu02} showed how these two techniques can be used to study the electronic structure of individual grains of \MB; here we apply them to quantify the changes as a function of either Al- or C doping. In particular, orientation resolved EELS will allow us to distinguish the behavior of the \PI\ and \Sig\ band for the different doping concentrations.
The samples of \MAB\ and \MBC\ were previously characterized for their superconducting properties;\cite{Zambano05, Cooley05PRL} we will correlate the changes in the measured electronic structure with these properties. Moreover, we use first-principles calculations to compare the changes in the EELS spectra to those in density of unoccupied B $p$ states for the different doping concentrations, and explore the filling of the \Sig\ band close to the Fermi-level.

\section*{Experimental Setup}
The results presented in this paper were obtained using
a JEOL3000F transmission electron microscope (TEM), equipped with an ultra high resolution
(UHR) objective lens pole piece, a Fischione high-angle annular dark-field (HAADF)
detector , a post column Gatan imaging filter (GIF), and a NORAN
X-ray detector. The instrument is capable of being operated in
either the conventional TEM or scanning TEM (STEM) mode.  For the atomic
resolution imaging and analysis results shown here, the incoherent HAADF imaging mode (or Z-contrast imaging) in the STEM was
used exclusively. The key to atomic resolution in
STEM microscopy\cite{james99-UM, james98-EM, nellist99, Klie05Micron} is the formation of
the smallest possible electron probe with sufficient probe current
(40 pA) to obtain statistically significant images and spectra.
The electron probe is optimized using the electron ``Ronchigram,''
or ``shadow image'' (for a more detailed description see
Refs.~\onlinecite{james99-UM}, \onlinecite{cowley86}), to obtain a
probe-size of $\sim$ 1.5 \AA\ for these experiments.

The experimental setup of this microscope allows  the
low-angle scattered electrons to be used, which do not contribute to the
Z-contrast image, for electron energy-loss spectroscopy (EELS).\cite{egerton86} As the two techniques
do not interfere, this means that Z-contrast images can be used to position the electron probe at the desired spot on the sample while acquiring spectra.\cite{james99-UM, fertig81, browning93}  The lens conditions in the microscope and spectrometer were set up for the smallest probe size, with a convergence angle ($\alpha$) of 11 mrad, a detector inner angle of 30 mrad, and an EELS spectrometer collection angle ($\theta_c$) of 25 mrad.

The
physical principle behind EELS relates the interaction of fast electrons with the sample to cause either collective
excitations of electrons in the conduction bands (plasmons) or discrete transitions between atomic energy levels.\cite{egerton86} In this study we will focus on the the latter class, using boron (B) $K$ edge core-loss spectra, which is comprised of transitions from the B $1s$-states into the unoccupied B $2p$-states above the Fermi-energy ($E_F$). In the experiment performed here, EEL spectra of the B $K$ edge are acquired directly from grains in the [110] and [001] orientations with an acquisition time of 3 s per spectrum. The experimental spectra shown here are a sum of 15 individual spectra, added to increase the signal-to-noise ratio of the near-edge fine structure. Further,  unless stated otherwise the background is subtracted from each spectrum and the resulting data are further deconvoluted with the zero-loss peak to remove the effects of plural scattering from the core-loss spectra,\cite{egerton86}.

The simulated EELS spectra shown in this paper are calculated using first-principles methods based on density-functional theory (DFT).\cite{Hohenberg65, Parr89} We used the newly updated \texttt{TELNES.2} package included in the \texttt{WIEN2K} code,\cite{Blaha01} a full-potential linear augmented plane-wave (FLAPW) plus local-orbitals method within DFT.  The generalized gradient approximation (GGA) proposed by Perdew, Burke and Ernzerhof \cite{Perdew96} was used for the exchange correlation. Muffin-tin radii ($r_{MT}$) were 2.00 bohr for Mg, and 1.65 bohr for B, and $r_{MT} k_{max}$  was taken to be 7.0.  The angular momentum ($l$) expansion up to $l_{max}$  for the potential and charge density representations was used in the calculations.
At convergence, the integrated difference between input and output charge densities was less than $10^{-4}$.  For all structures, 252 $k$-points in the irreducible Brillouin zone were used in the calculations. We used the experimental lattice parameters, as given in Table \ref{Tab:Samples}, in our calculations of the EELS fine-structure for the different doping concentrations.

\section*{Sample Preparation}

The \MABC\ samples for this study where prepared using various reaction and sintering conditions to create material with a homogeneous distribution of the doping element, minimize secondary phases such as MgB$_4$, MgO, AlB$_x$ or MgC$_x$,  and to maximize the grain size. Here, we will only briefly describe the different sintering conditions for the various samples, while more detailed descriptions can be found elsewhere.\cite{Wilke04, Zambano05, Cooley05PRL}
Undoped \MB\ and \MAB\ samples were prepared in a high-temperature process that lasted for nearly 96 hrs,with thorough annealing to obtain homogeneity. Initially, the appropriate mixture of Mg, B, and Al powders were heated to 1200\DEG C and kept at this temperature for 1 h. The temperature was then decreased to 700\DEG C at a rate of $0.1^{\circ}C\  min^{-1}$ and kept at 700\DEG C for 5 hrs. Finally, the sample was brought to room temperature at a rate of $20^{\circ} C\ min^{-1}$.\cite{Zambano05} The resulting materials have an average grain-size of $>~1~\mu m$ and do not show show a significant number of impurity or secondary phases.

The \MBCa\ sample was prepared from a  carbon-doped B fiber by Ames Laboratory.\cite{Wilke04} The \MBCb\ sample was made by mixing the appropriate ratio of Mg and B$_4$C powders and subsequently heating it to 950\DEG C for 1 h followed by slow cooling. X-ray diffraction analysis of this material showed the presence of a small concentration of Mg$_2$C$_3$ impurity phases.

\begin{figure}[h]
\begin{center}
\includegraphics[height=60mm]{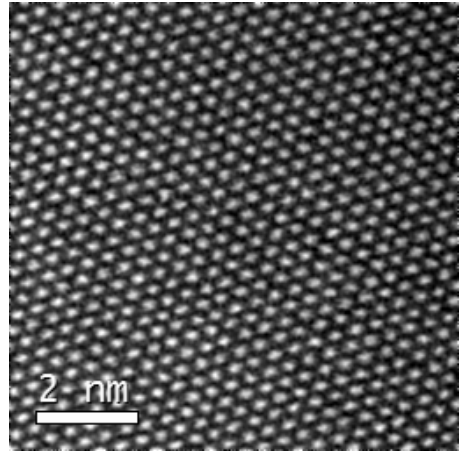}
\caption{High-resolution Z-contrast of 25\% Al-doped \MB\ [001] with no apparent superstructure due to Al-ordering.}
\label{Fig:Z-contrast}\end{center}
\end{figure}

For all the samples studied here, we found that the doping concentration within the grains that were studied appears to be homogeneous, and no ordering of the dopants was found in any sample. We found that the precise doping concentration varies slightly from grain to grain, in particular, in those samples with shorter sintering time. We used EDX and EELS analyses to determine the doping concentration of the individual grains probed, and the respective doping concentration given for each sample in this paper applies only to the individual grains that were studied. Figure \ref{Fig:Z-contrast} shows an example of a typical high-resolution Z-contrast image of a grain in the [001] projection found in \MAcB. Similar images can be obtained for all other doping concentrations. The bright spots in this image show the Mg/Al-columns in this projection of the \MB\ structure. Due to the small scattering amplitude at large scattering angles, the B-columns are not visible in this image. Since the image intensity is proportional to $Z^{1.7}$, and Al ($Z$=13) is slightly heavier than Mg ($Z$=12), while C ($Z$=6) is slightly heavier than B (Z=5), ordering or clustering of of the Al- or C atoms would be visible in such Z-contrast images as variations of the atomic column contrast. For instance, oxygen ordering in grains of \MB\ on the B sublattice was previously observed by similar Z-contrast imaging.\cite{klie02APL} This makes it likely that ordering or clustering of dopants should also be visible here. From Figure \ref{Fig:Z-contrast}, which is shown here as a representation of all other doping concentrations, it can be clearly seen that no superstructure due to dopant ordering can be found. A summary of the properties of the different sample materials used in this study is given in Table \ref{Tab:Samples}.

\section*{Results}

\begin{figure}[h]
\begin{center}
\subfigure[] {
\includegraphics[width=80mm]{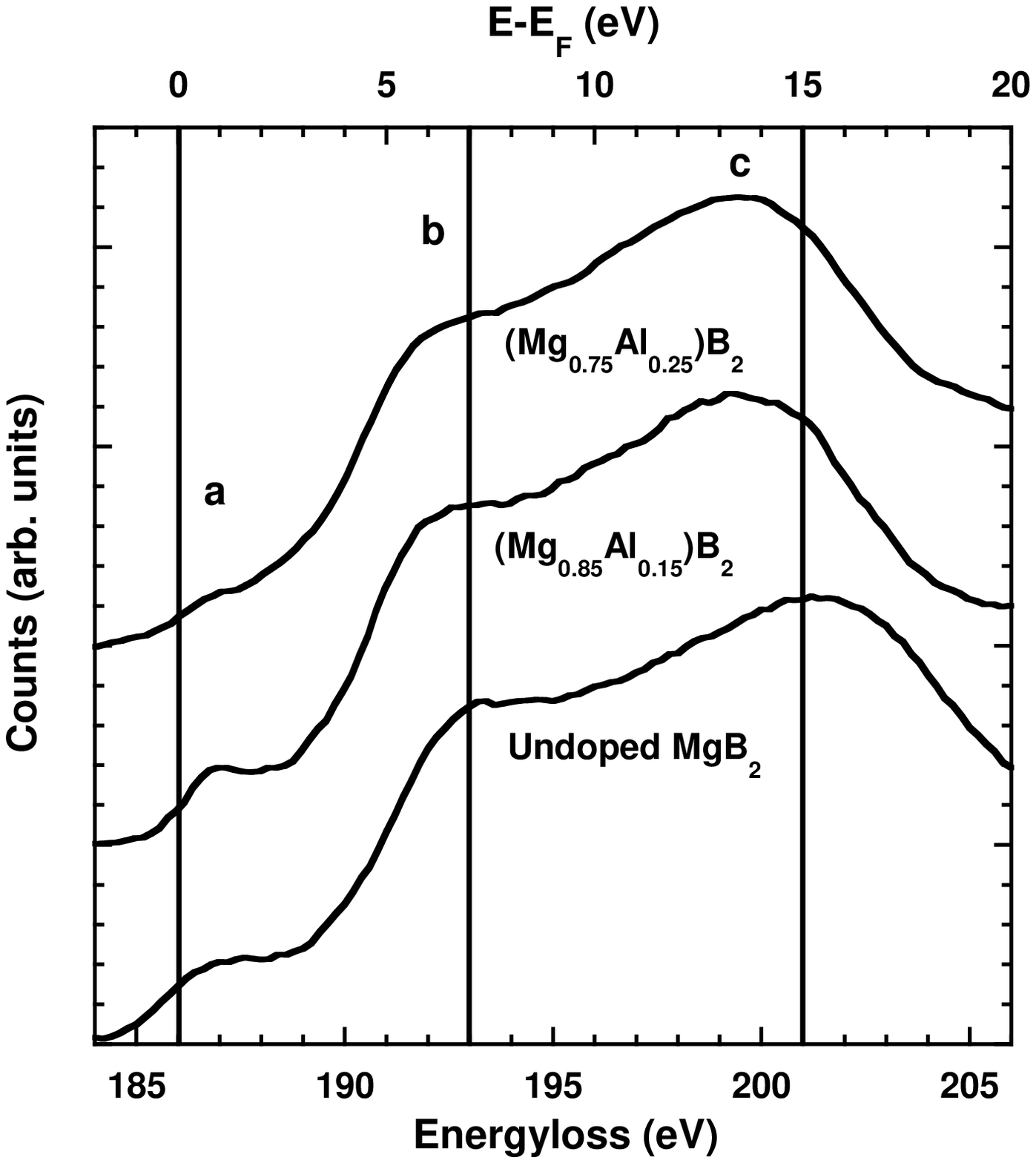}
\label{Fig:Al001-exp}
}

\subfigure[] {
\includegraphics[width=80mm]{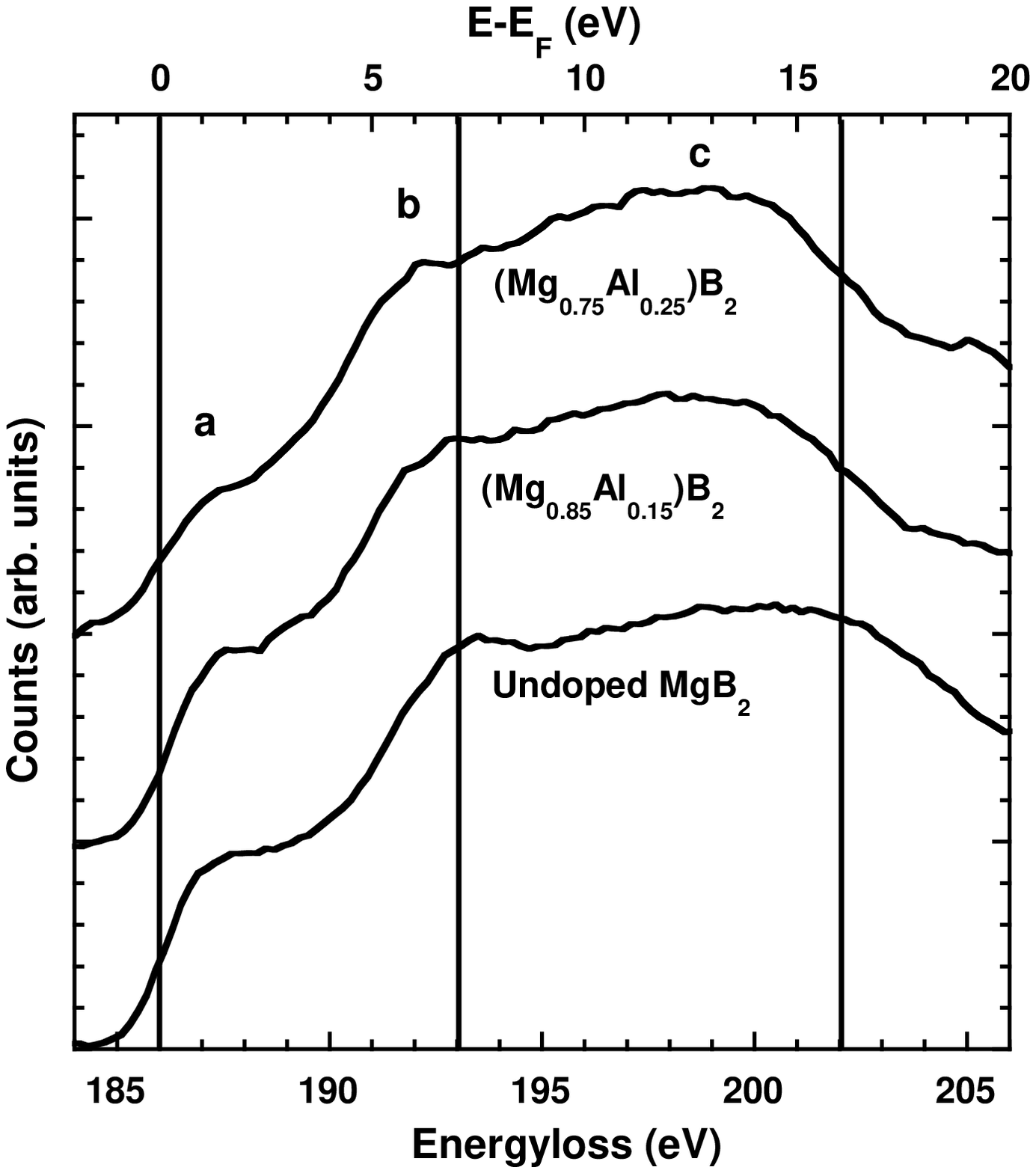}
\label{Fig:Al110-ext}}

\caption{EELS of B $K$ edge for different Al-doping concentrations from grains in the a) [001] and b) [110] orientation. }
\end{center}
\end{figure}

Figures \ref{Fig:Al001-exp} and \ref{Fig:Al110-ext} shows the B $K$ edge core-loss EELS spectra for the different Al-doping concentrations.
The spectra are aligned at the edge onset for undoped \MB, which is located at 186.0 eV and are further
offset in the vertical direction to clearly show the differences in the near-edge fine-structure for the different doping concentrations.
In this projection the experimental B $K$ edge spectrum contains three major features, or peaks within the first 20 eV above the edge onset. The small shoulder, or pre-peak intensity is labeled throughout this paper as peak \textit{a}; the pre-peak is followed by a shoulder of higher intensity at $\sim$ 193 eV in pristine \MB\ and is labeled peak \textit{b}. The main feature of the B $K$ edge, which is located at 202 eV, represents a broad peak without any further fine-structure and is labeled peak \textit{c}.
We have added three vertical lines to all the figures containing EELS spectra, at the edge onset (186 eV), the position of peak \textit{b} (193 eV) and the broader peak \textit{c} at 202 eV in pristine \MB. Figure \ref{Fig:Al001-exp} shows the spectra taken from \MAB\ grains in the [001] projection.

With increasing Al concentration, it can be clearly seen from Figure \ref{Fig:Al001-exp} that peaks \textit{b} and \textit{c} are shifting towards lower energies. In addition, the peak intensities change for the different doping concentrations: the pre-edge peak  intensity (\textit{a}) decreases significantly for \MAbB\ and is nearly completely vanished for \MAcB. Further, the intensity of peak \textit{c} seems to be increased for both \MAbB\ and \MAcB.

The spectra from grains in the [110] orientation (Figure \ref{Fig:Al110-ext}) are very different from those of the [001] orientation. The most obvious difference is the higher pre-peak intensity (peak \textit{a}) in the [110] orientation compared to the the [001] orientation, and we have previously discussed these differences in the context of the two bands of the boron $p$ states in \MB.\cite{klie03-MgB2, klie03} However, the spectra in the [110] orientation show  similar trends as seen in the [001] orientation, in particular the energy shifts of peaks \textit{b} and \textit{c} as well as the decrease in the intensity of the pre-edge peak \textit{a}.
While this pre-edge peak intensity decreases as a function of Al doping, it is still clearly visible for \MAcB.
In undoped \MB, peak \textit{b} is clearly visible, while  for \MAcB\ it is barely noticeable as a shoulder at $\sim$ 191.8 eV.

\begin{figure}[h]
\begin{center}
\subfigure[] {
\includegraphics[width=80mm]{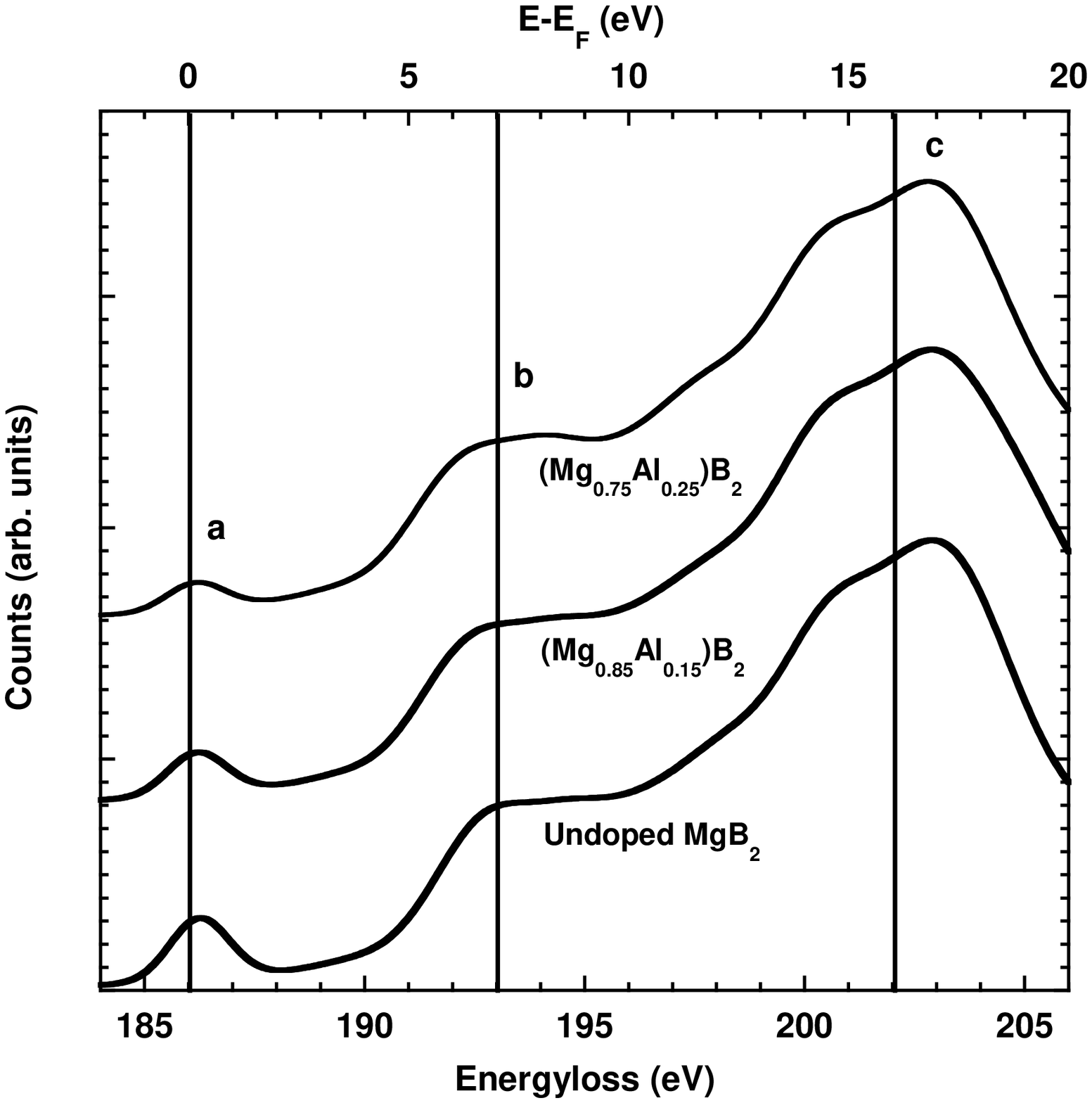}
\label{Fig:Al001-dft}
}

\subfigure[] {
\includegraphics[width=80mm]{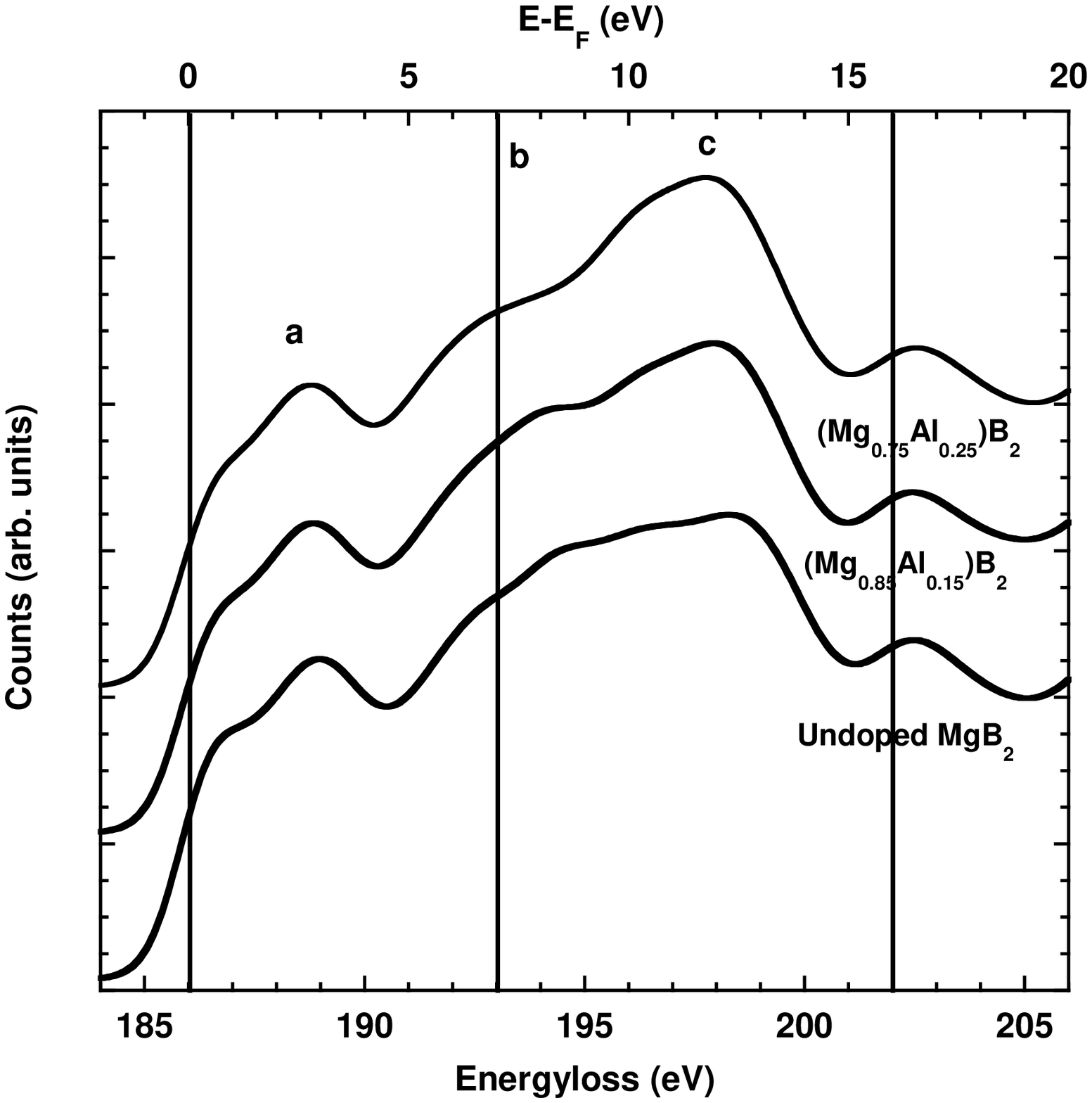}
\label{Fig:Al110-dft}
}

\caption{Calculated B $K$ edge for different Al-doping concentrations using DFT with VCA for grains in the a) [001] and b) [110] orientation. }\end{center}
\end{figure}

Figure \ref{Fig:Al001-dft} and \ref{Fig:Al110-ext} show the calculated EELS spectra for \MAB\ in the [001] and [110] orientation using the virtual crystal approximation (VCA). Unlike previously published spectra, these calculations take into consideration the experimental unit-cell volume which might be considerably different from the calculated ones, and this results in significant changes of the near-edge fine structure of the calculated spectra.
The theoretical spectra shown here are smoothed by a Gaussian function with 0.8 eV width at half-maximum to simulate the instrument resolution.
The simulated EELS spectrum for undoped \MB\ shows good agreement with the experimental results, in that the positions of peaks \textit{a} and \textit{b} are reproduced accurately, as well as decrease in intensity of peak \textit{a} for increasing Al concentration.  For the [001] orientation (Figure \ref{Fig:Al001-dft}), the energy-position of peak \textit{c} is slightly higher in the calculated spectra, and the energy shift of peaks \textit{b} and \textit{c} appears to be slightly smaller than in the experimental spectra. Figure \ref{Fig:Al110-dft} shows more structure of the pre-peak \textit{a} than seen in the experimental spectra, and peak \textit{b} is not distinguishable from the broader shoulder that appears at 198.5 eV  in the calculated spectrum of pristine \MB. This peak is not visible in the experimental spectra, and the intensity of peak \textit{c} at 201.5 eV is underestimated by the theoretical spectra.
However, the peak at 198.5 eV becomes stronger and more distinct with increasing Al concentration, while no changes can be seen for the peak at 201.5 eV.

The B $K$ edge EELS spectra for C doped \MB\ are shown in Figure \ref{Fig:C001-exp} and \ref{Fig:C110-ext}, where undoped \MB\ is compared to \MBCa\ and \MBCb\ in the two crystal orientations, [001] and [110]. Similar to the Al doped samples, the B $K$ edge in C-doped \MB\ shows a measurable decrease in the peak intensity of the pre-peak \textit{a} in both crystal orientations, and a small shift of peak \textit{b} in the spectra from \MBC\ grains as the C concentration is increased. While peak \textit{c} shifts downwards in energy for \MBCb\, there is no noticeable shift of peak \textit{c} in \MBCa. However, with increasing C doping concentration one can observe an increase in the intensity of peak \textit{c}, which is located at 202 eV in pristine \MB. The spectra from grains in the [110] orientation (Figure \ref{Fig:C110-ext}) show a shift in energy of peak \textit{b}, but no measurable shift of peak \textit{c} as a function of C doping. In addition to this shift, peak \textit{b} (at 193.5 eV in \MB) becomes more distinguishable for higher C concentrations, a similar trend can be seen for the broader peak \textit{c}.

\begin{figure}[h]
\begin{center}
\subfigure[]
{
\includegraphics[width=80mm]{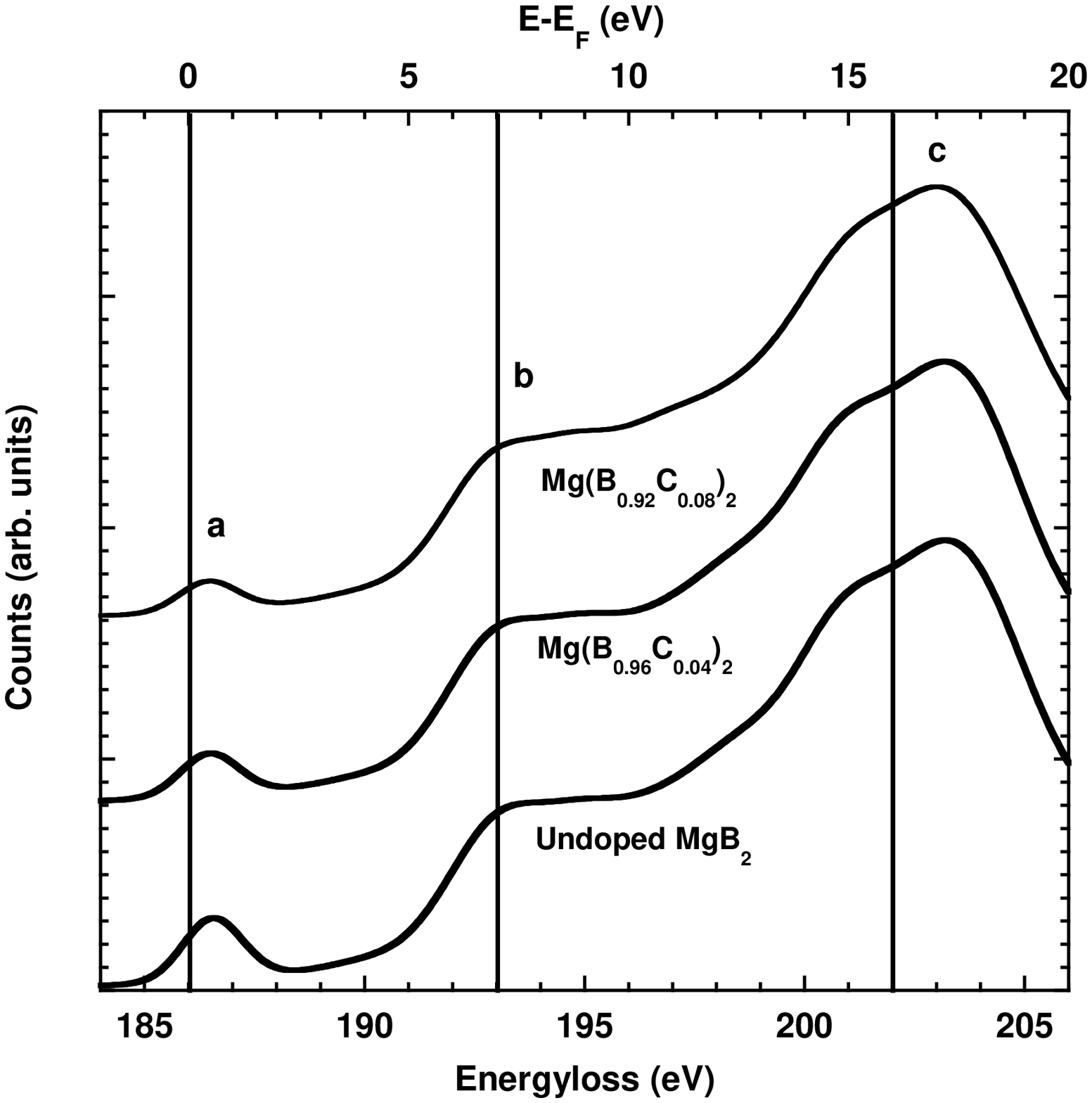}
\label{Fig:C001-dft}
}

\subfigure[]
{
\includegraphics[width=80mm]{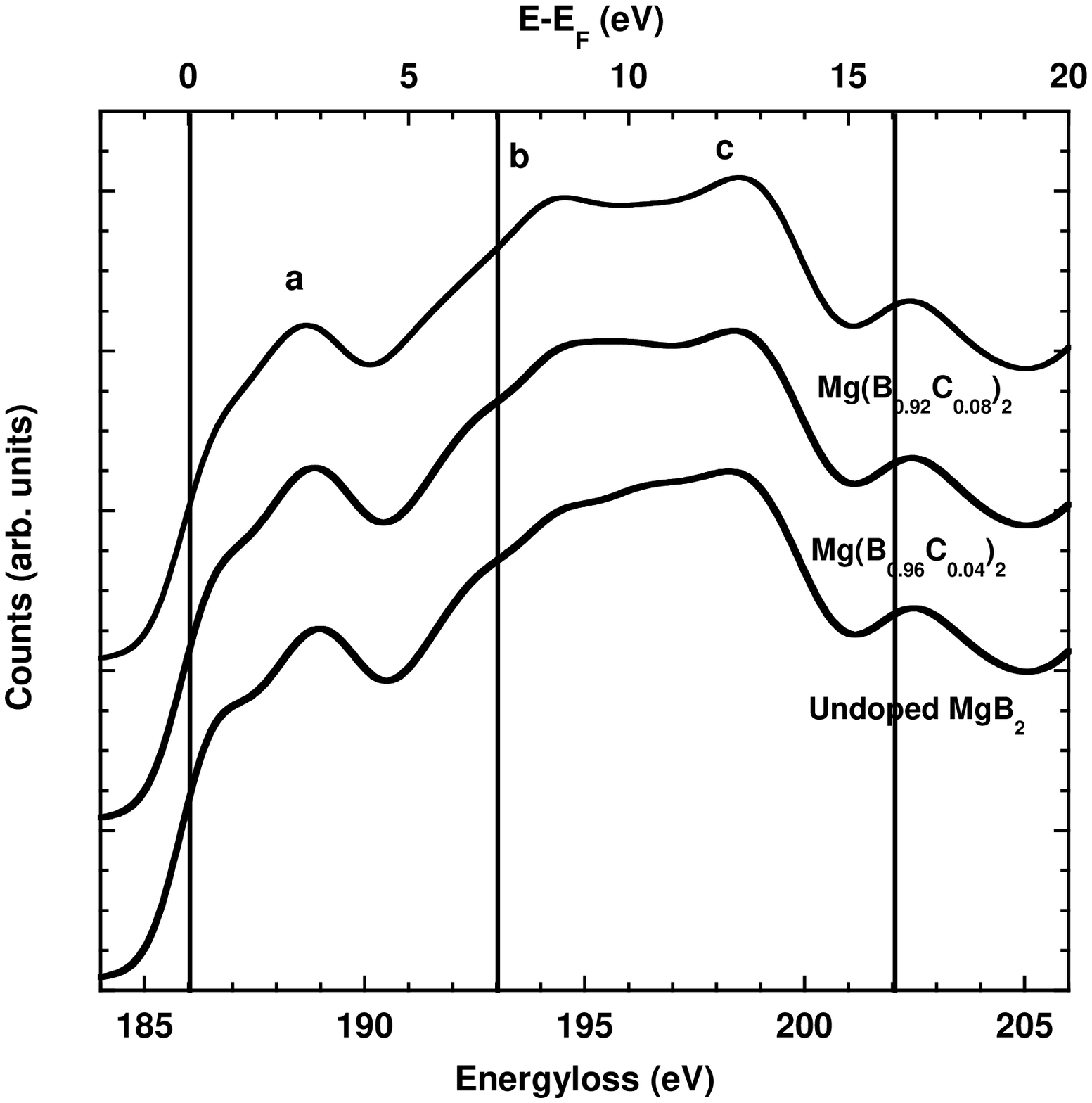}
\label{Fig:C110-dft}
}

\caption{Calculated B $K$ edge for different C-doping concentrations using DFT with VCA for grains in the a) [001] and b) [110] orientation.}
\end{center}
\end{figure}

The results of the DFT calculations for C doped \MB\ in the [001] and [110] orientation, shown in Figure \ref{Fig:C001-dft} and \ref{Fig:C110-dft}, are in general agreement with the experimental spectra. In calculation these spectra, the VCA was used and the resulting spectra were broadened by 0.8 eV to match the experimental energy resolution. Figure \ref{Fig:C001-dft} shows the decrease in the pre-peak intensity (peak \textit{a}) with increasing C concentration and a small shift of peak \textit{c} towards lower energy for \MBCb. However, the calculated spectra show an intensity of peak \textit{b} that is significantly lower than measured experimentally. The calculations for the [110] orientations (Figure \ref{Fig:C110-dft}) show a decrease in the pre-peak intensity, and reflect the positions of peak \textit{b} adequately. However, the neither the position of peaks \textit{c} nor the shape and intensity of peak \textit{b} are not reproduced in the calculated spectra. Similar to Fig.~\ref{Fig:Al110-dft}, the calculated spectra show an additional intensity at 198.5 eV, while underestimating the intensity of peak \textit{c}. Further, the peak at 198.5 eV increases in intensity as the C doping concentration is increased.

While many of the general changes in the first 15-20 eV of the B $K$ edge fine structure are well reproduced by the DFT calculations, the intensity of peak \textit{b} at $\sim$ 193 eV  and peak \textit{c} in teh [110] orientation are only poorly reflected in the theoretical spectra. We have previously shown that the peak at193 eV is correlated to oxygen rich areas in \MB\, or even amorphous $BO_x$ principates or surfaces layers.\cite{zhu02, Idrobo05, klie02APL} Since we could not observe any superlattice structure in the \MB\ grains that would indicate O inside the \MB\ grain, it is reasonable to assume that the exposure of the samples to air has caused the formation of an amorphous boron-oxide layer at the top and bottom surfaces. The shift of this peak at $\sim$ 193.0 eV is an artifact of the post-acquisition energy-scale alignment, and further indicates a shift in the Fermi-level upon doping \MB\ with electron.

\section*{Discussion}

\begin{figure}[h]
\begin{center}
\subfigure[] {
\includegraphics[width=80mm]{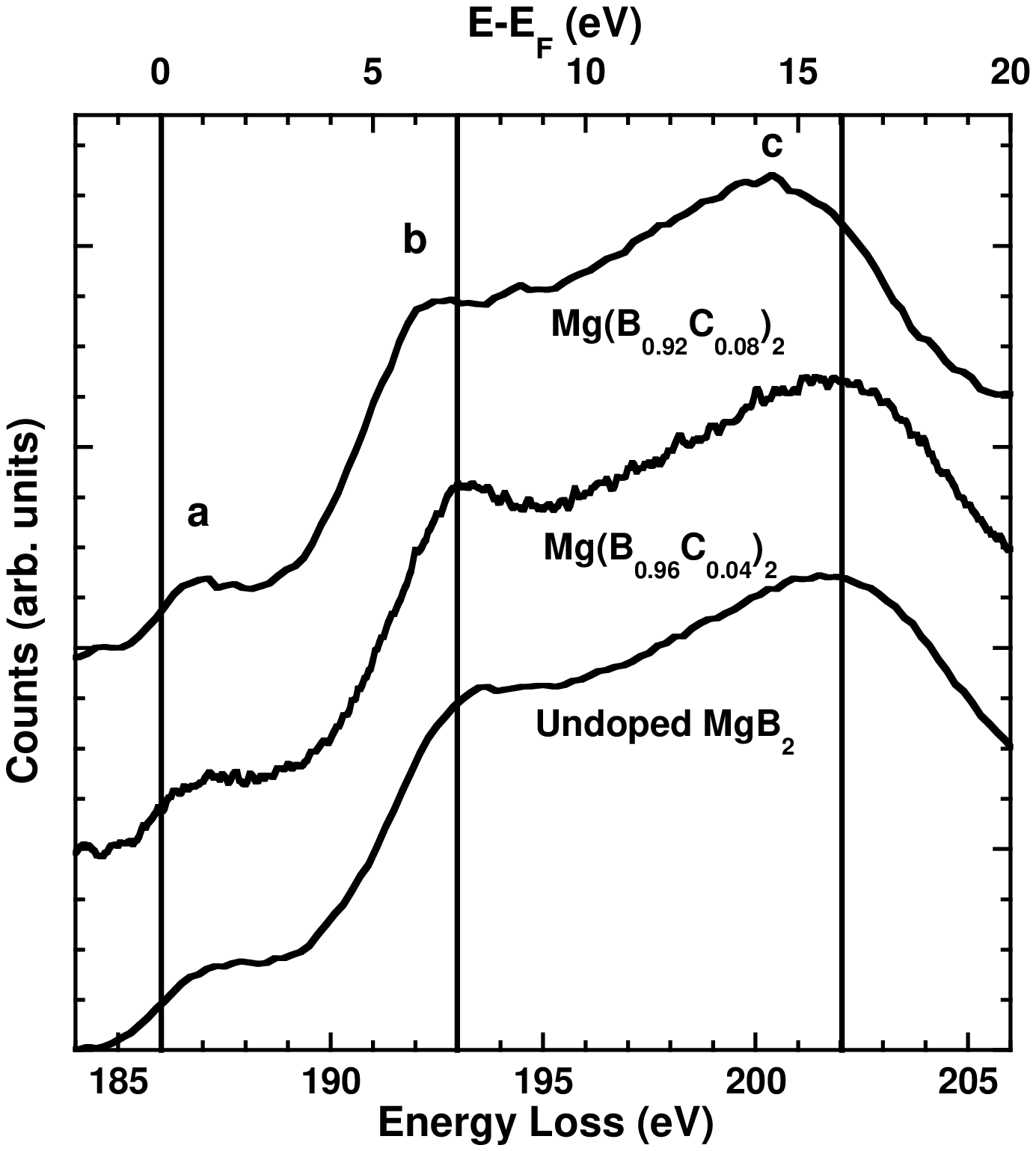}
\label{Fig:C001-exp}
}

\subfigure[] {
\includegraphics[width=80mm]{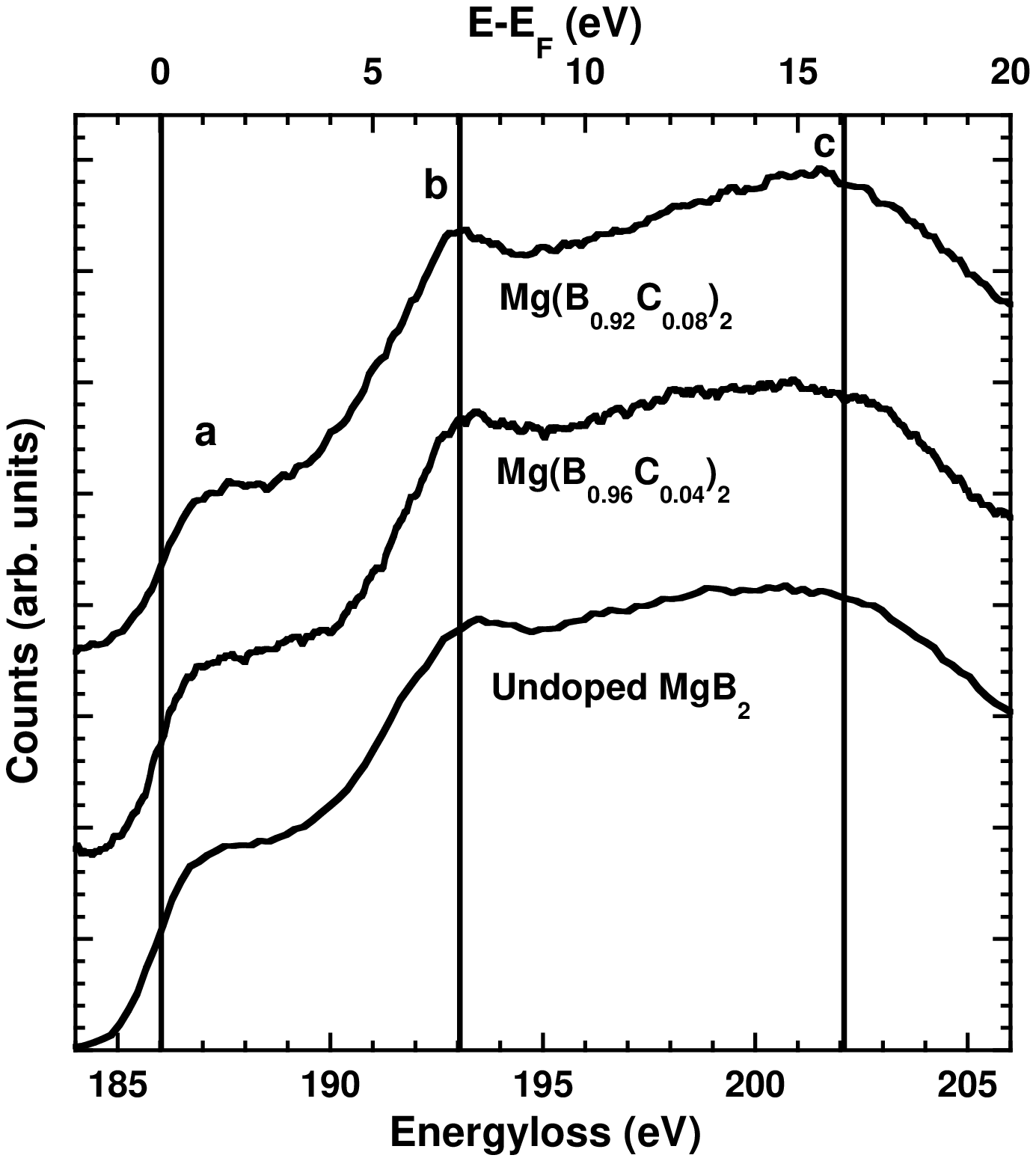}
\label{Fig:C110-ext}}

\caption{EELS of B $K$ edge for different C-doping concentrations from grains in the a) [001] and b) [110] orientation}
\end{center}
\end{figure}

In order to interpret the changes in the B $K$ edge fine-structure as a function of Al and C doping, a complete understanding of the orientation dependance of the B $K$ edge is needed.
 It has been shown previously that the B $K$ edge near-edge fine-structure of \MB\ will change significantly as a function of grain orientation with respect to the incoming electron probe (as shown in Figure \ref{Fig:Al001-exp}, \ref{Fig:Al110-ext} or Figures \ref{Fig:C001-exp} and \ref{Fig:C110-ext}) due to the fact that \MB\ is an anisotropic material.\cite{zhu02, klie03-MgB2, klie03} Since \MB\ is a layered compound with graphite-like layer of B atoms  and intercalated Mg layers, the orientation of the electron probe with respect to these layer can excite different transitions as the angle between the electron probe and the 2-dimensional B layer changes. We have previously shown that two crystal orientations, [001] and [110] (i.e.~perpendicular and parallel to the hexagonal B-layers) are sufficient to separate the contributions of the boron \Sig\ and the \PI\ bands.\cite{klie03-MgB2, klie03} For EELS core-loss spectra that have been acquired from \MB\ grains in the [001] projection, we have shown the majority of the spectral contribution  ($\sim 90\%$) stems from the transitions perpendicular to the incoming beam direction, i.e. promoting the B 1s-electron into the unoccupied \Sig\ band. For grains in the [110] orientation, only  half the observed intensity comes from those \Sig\ band transitions, the other half is made up by transitions into the \PI-band. This can be understood by considering the fact that we use a highly convergent electron probe and a large spectrometer acceptance angle, which means that there is significant momentum transfer into transition perpendicular to the incoming beam direction.\cite{klie03}

\begin{figure}[h]
\begin{center}
\includegraphics[width=80mm]{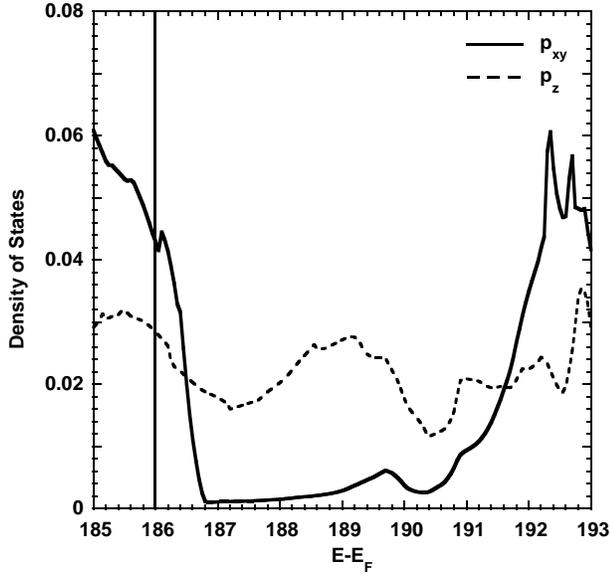}
\caption{Density of boron $p$-states in the vicinity of the Fermi-level. The Fermi-energy ($E_F$) is shown at 186 eV, while the two separate bands are shown between -1.0 eV and 7.0 eV with respect to the Fermi energy.}

\label{Fig:dft}

\end{center}
\end{figure}

Figure \ref{Fig:dft} shows the density of states (DOS) in the vicinity of $E_F$ as computed by DFT for undoped \MB. The unoccupied \Sig\ states have high intensity at $E_F$ that drops to zero within 0.8 eV above $E_F$. The DOS of the \Sig-band then remains low up to $\sim$5 eV above $E_F$ and finally increases to form the \Ss-peak at 192.0 eV.
Meanwhile, the \PI\ states have a nearly flat distribution of the DOS as the energy increases from 186 eV to 193 eV, with the exception of the small peak at 189 eV.
As described above, any experimental EELS spectrum will contain contribution from both of those bands,\cite{klie03, Brydson89} which means that the pre-peak in the experimental spectra contains transitions from the B $1s$ states into the sum of the high density of unoccupied \Sig\ and the flat \PI\ states at $E_F$, while peak \textit{b} contains transitions into the \Ss\ at $\sim$ 192 eV. The changes in the B $K$ edge fine-structure for the different crystal orientations can be best understood by a varying mixture of the \Sig\ and the \PI\ states to the total B $K$ edge spectrum.

The superconducting properties of \MB, in particular $T_c$,  are strongly correlated to the high density of \Sig-states close to the Fermi-level, and the the dominance of these transitions in the spectra from the [001] orientation gives us a unique experimental window, through which the electronic structure important for superconductivity can be accessed. It is widely believed that filling of these \Sig-states by electron doping will decrease the density of superconducting carriers (which are holes), and thus $T_c$ should decrease. The filling of the empty $p$ states will results in an increase in the Fermi level, and one would expect that $T_c = 0\ K$ once all the \Sig-states are filled (i.e. the Fermi-level shifted by more than 0.8 eV) This shift in the Fermi level should result in a chemical shift of the B $K$ edge as well as a decrease in the pre-edge peak intensity.
Indeed, this is exactly what is seen in Figs.~\ref{Fig:Al001-exp} and \ref{Fig:C001-exp}.

\begin{figure}[h]
\begin{center}
\subfigure[]
{
\includegraphics[width=80mm]{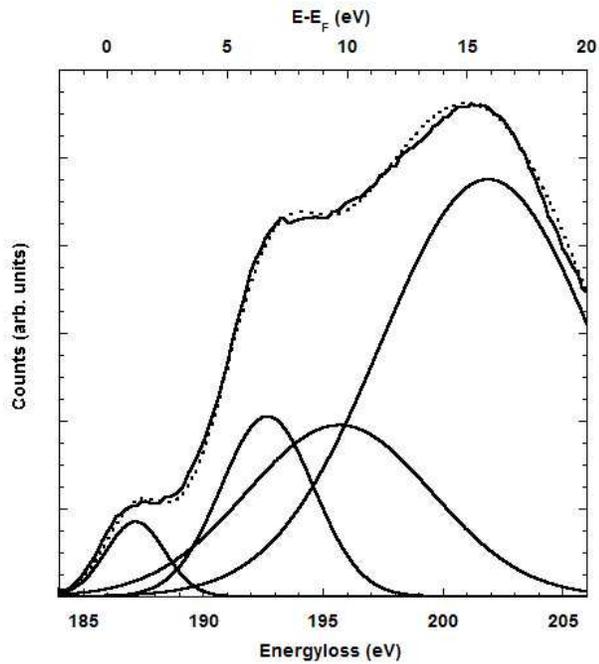}
\label{Fig:fit}
}

\subfigure[]
{
\includegraphics[width=80mm]{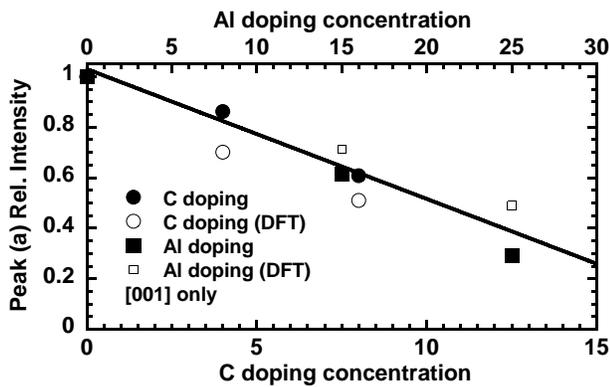}
\label{Fig:Prepeak vs doping}
}
\caption{a) Example of the four-Gaussian fit (dotted line) with the experimental data (solid line) for undoped \MB, used to calculate the decrease in pre-peak intensity relative the \Ss-peak as a function of electron doping. b) Relative pre-peak intensity as a function of electron-doping as a result of the four-Gaussian fit. The open symbols represent the theoretical and the full symbols the experimental results; the least-square fit is shown as a solid line.}

\end{center}
\end{figure}

To quantify the changes in the B $K$ edge, we will use the spectra from the [001] orientation only, since they provide us with the majority of \Sig-states transitions, and either the chemical shift or the pre-peak intensity can be used to correlate the hole-state density to the superconducting properties.
However, since the energy-resolution of the experimental EELS spectra is only $\sim$1.0 eV  and a chemical shift of 0.8 eV is sufficient to fill most of the available \Sig-states, we will use the pre-peak intensity as a more reliable measure.
A Gaussian function was used to fit the pre-edge peak at 0-1 eV above $E_F$, three more Gaussian functions were then used to fit the remaining peaks of the B $K$ edge.
Figure \ref{Fig:fit} shows the spectrum for undoped \MB\ and a fit of the experimental data using four Gaussians of varying width and intensity. The pre-peak intensity of the experimental spectrum consists of a Gaussian located at 186 eV, while the \Ss\ peak is formed by one Gaussian at 192 eV and a broader Gaussian at 195 eV. Similarly, we have fitted all the experimental spectra of the different Al- and C doping concentrations shown in Figures \ref{Fig:Al001-exp} and \ref{Fig:C001-exp}. Figure \ref{Fig:Prepeak vs doping} shows the relative pre-peak intensity as calculated from the Gaussian fits for the different Al- and C-doping concentrations. The pre-peak intensity is normalized to the fitted \Ss-peak intensity to account for variations in sample thickness or acquisition times. Similarly, we have fitted four Gaussian functions to all the calculated spectra shown in Figs.~\ref{Fig:Al001-dft} and \ref{Fig:C001-dft}. We can clearly see that within the experimental uncertainty the pre-peak intensities for both Al and C-doped decrease linearly with increasing doping concentration, and the least square fit is plotted in Fig.~\ref{Fig:Prepeak vs doping}.
Thus, for the first time there is direct spectroscopic evidence that band filling occurs as \MB\ is doped with electrons.
Extrapolating the linear fit in Fig.~\ref{Fig:Prepeak vs doping}, one will find that relative pre-peak intensity will vanish completely at $x_{crit}^{EELS}\sim$ 40\% Al-doping or $y_{crit}^{EELS}\sim$20\% C . This means that all the empty \Sig-states close to $E_F$ will be filled completely at $x >$ 40\% Al-doping, thus resulting in a collapse of the high-temperature 2-band superconductivity in \MAB.\cite{Cooley05PRL} It has been previously reported that $x_{crit}$ (the Al-composition at which $T_c\ \rightarrow 0\ K$)is at $\sim$ 33 - 40\%. However, $y_{crit}$ (the C-composition at which $T_c\ \rightarrow 0\ K$) has been measured to be $\sim$ 12 - 15 \%.\cite{Kortus05} This means that within the experimental uncertainties $x_{crit}^{obs} \approx\ x_{crit}^{EELS}$, but $y_{crit}^{obs} <\ y_{crit}^{EELS}$.

This means that, although the changes in the B $K$ edge spectra for Al and C-doped \MB\ appear to be very similar, there seems to be a fundamental difference in the mechanism that governs the measured loss of superconductivity in doped \MB\ for these dopants. The drop in $T_c$ with increasing Al concentration can be successfully explained in terms of band filling, while the simple band-filling picture appears to underestimate the decrease in $T_c$ in C-doped \MB. Similarly, Kortus et al.\cite{Kortus05} reported that the larger decrease of $T_c$ in C-doped \MB\ can be explained by an increased interband scattering rate, while the effect of Al-doping on $T_c$ is simply due to band-filling. Further, the authors ague that the two superconducting gaps in \MBC\ merge, resulting in an additional lowering of $T_c$ of about 6 K in \MB\ with 10\%-15\% C-doping.

\begin{figure}[h!]
\begin{center}
\subfigure[]
{
\includegraphics[width=80mm]{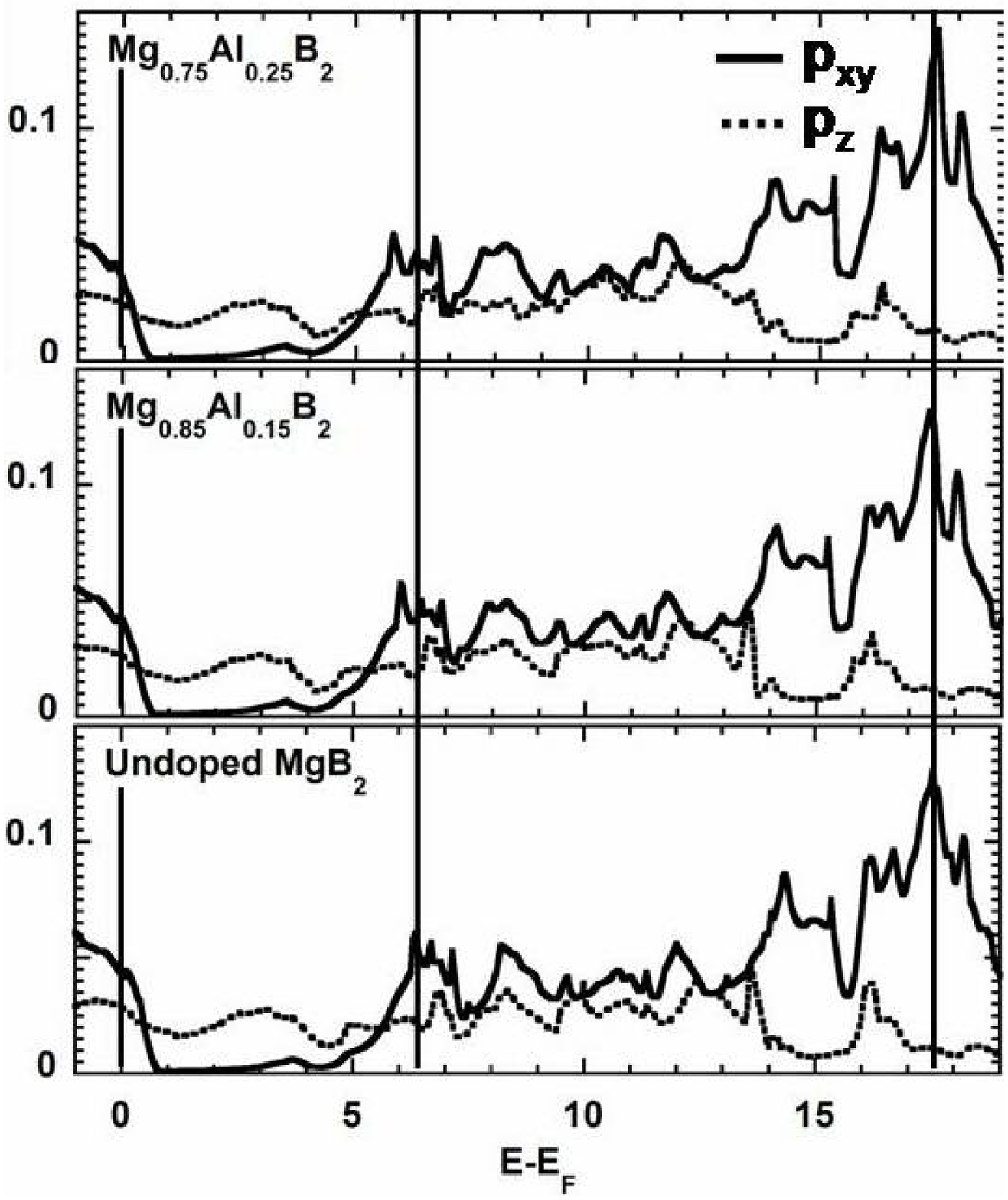}
\label{Fig:DOS-Al}
}

\vspace{20mm}

\subfigure[]
{
\includegraphics[width=80mm]{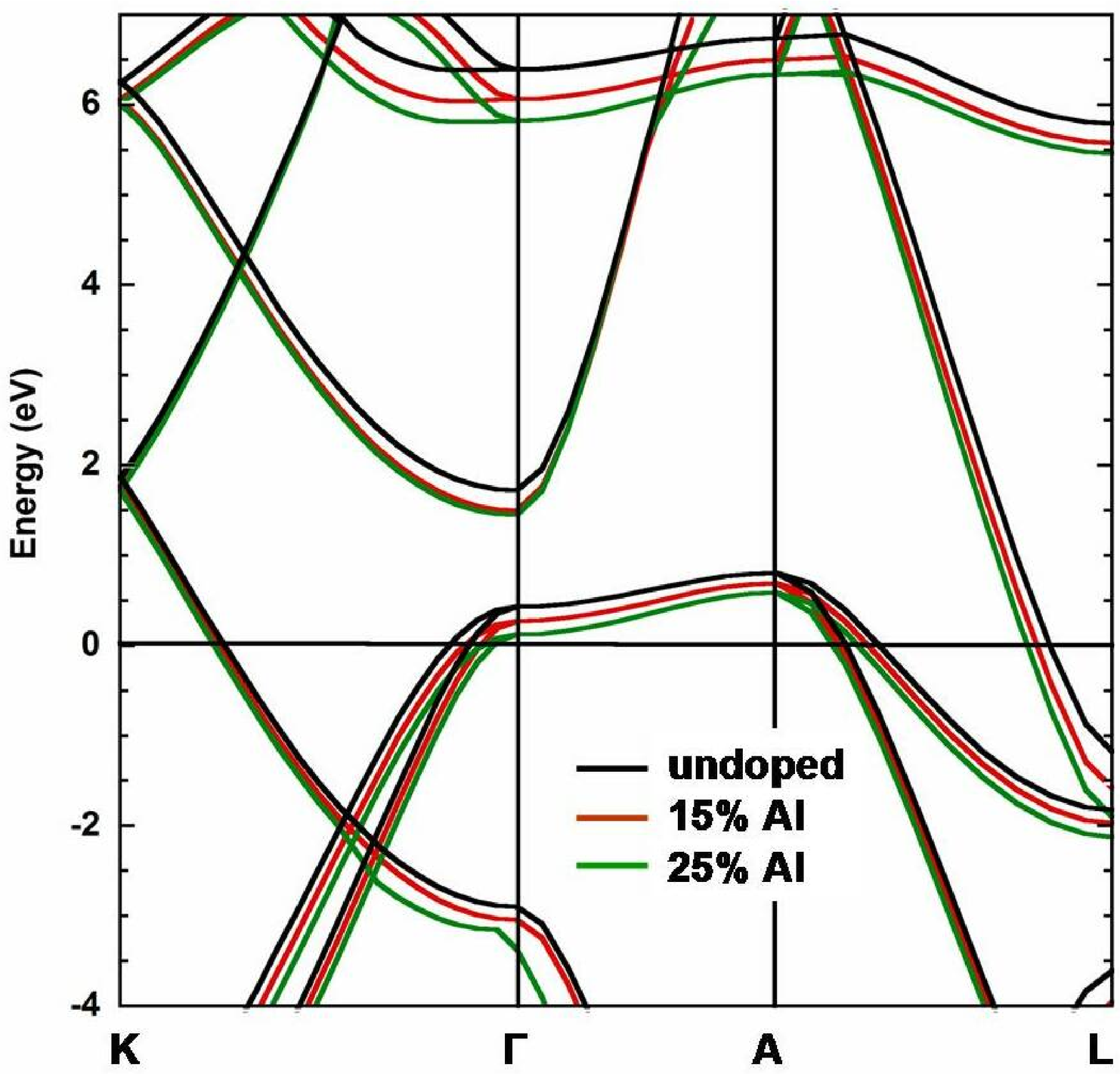}
\label{Fig:Bandstructure-Al}
}
\caption{\textbf{Color:} Calculated a) partial density of states and b) bandstructure of pristine and Al-doped \MB\  showing the \Sig-band shift for the different doping concentrations.}
\end{center}
\end{figure}

For a more in-depth look at the changes in the electronic structure of electron-doped \MB, we have calculated the density of states and the bandstructure for the compounds we have studied experimentally. Fig.~\ref{Fig:DOS-Al} shows the partial DOS of the \Sig\ and the \PI-band for \MB, \MAbB\ and \MAcB; Fig.~\ref{Fig:Bandstructure-Al} shows the bandstructure for the same compounds in the vicinity of the Fermi-level containing the \PI, the \Sig\ and the \Ss-bands.
The DOS of the \Sig-band for Al-doped \MB\ (Fig.~\ref{Fig:DOS-Al}) shows the decrease of the high density of states at the Fermi-level that is closely related to the measured pre-peak intensity change. Further, the \Ss-peak at $\sim$6 eV above $E_F$ shifts towards lower energies with increasing Al-doping concentration. The highest peak in this energy range located at 17.5 eV above $E_F$ (and corresponds to peak \textit{c} in Fig.~\ref{Fig:Al001-exp}) does not exhibit any visible shift, but increases in intensity as the Al-concentration is increased. The empty \PI-band states do not show any significant change for the different doping concentrations.

The band structure (Fig.~\ref{Fig:Bandstructure-Al}) reveals some further details related to the changes in the electronic structure upon Al-doping. Here, the high intensity at $E_F$ in the \Sig-band corresponds to the flat band close to the Fermi-level between $\Gamma$ and $A$, while the \Ss-peak corresponds to the flat band $\sim$ 6 eV above $E_F$. It can be clearly seen that with increasing Al-concentration the \Sig-band close to $E_F$ shifts lower in energy and at 25\% the band at $\Gamma$ is nearly completely submerged under the Fermi-level. The density of states at the $\Gamma$ point is particularly important, since it is these states at the Brillouin zone (BZ) center that couple to the $E_{2g}$ phonons and create the high $T_c$. Further, the \Ss-band appears to be shifting faster in energy than the bands close to $E_F$ as the Al-concentration increases, thus indicating a non-rigid shift of the \Sig\ bands applies. On the other hand, there appears to hardly any shift of the \PI-bands, which are located at $K$ in the BZ. Interestingly, this means that the extra electrons provided by the Al substituting for Mg fill only the \Sig\ hole-states rather than the \PI\ band. Therefore, this filling of the \Sig\ states, albeit non-rigid, can account for the decrease in $T_c$ that was previously reported. \cite{Cooley05PRL, Kortus05}

\begin{figure}[h!]
\begin{center}
\subfigure[]
{
\includegraphics[width=80mm]{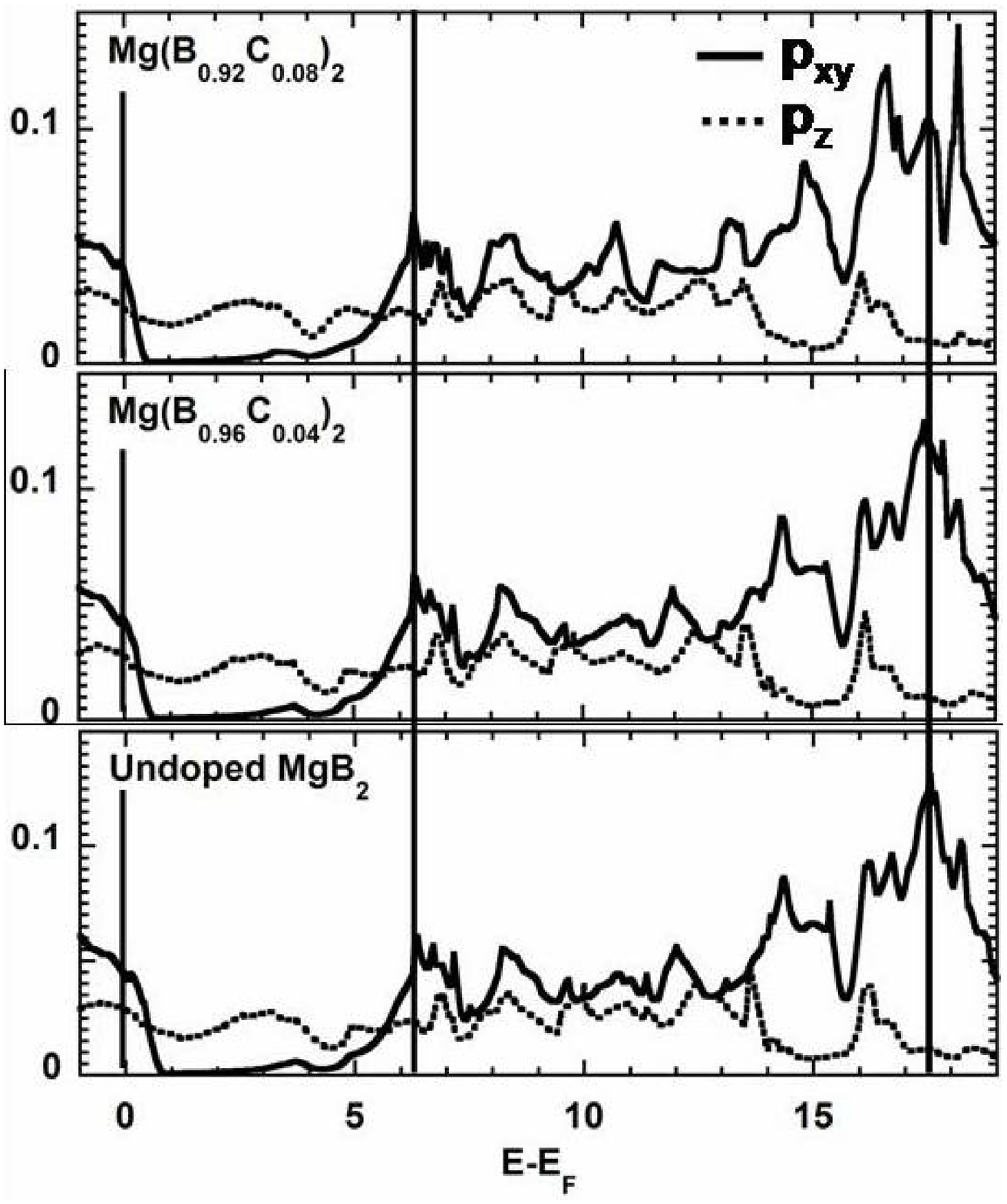}
\label{Fig:DOS-C}
}

\vspace{10mm}

\subfigure[]
{
\includegraphics[width=80mm]{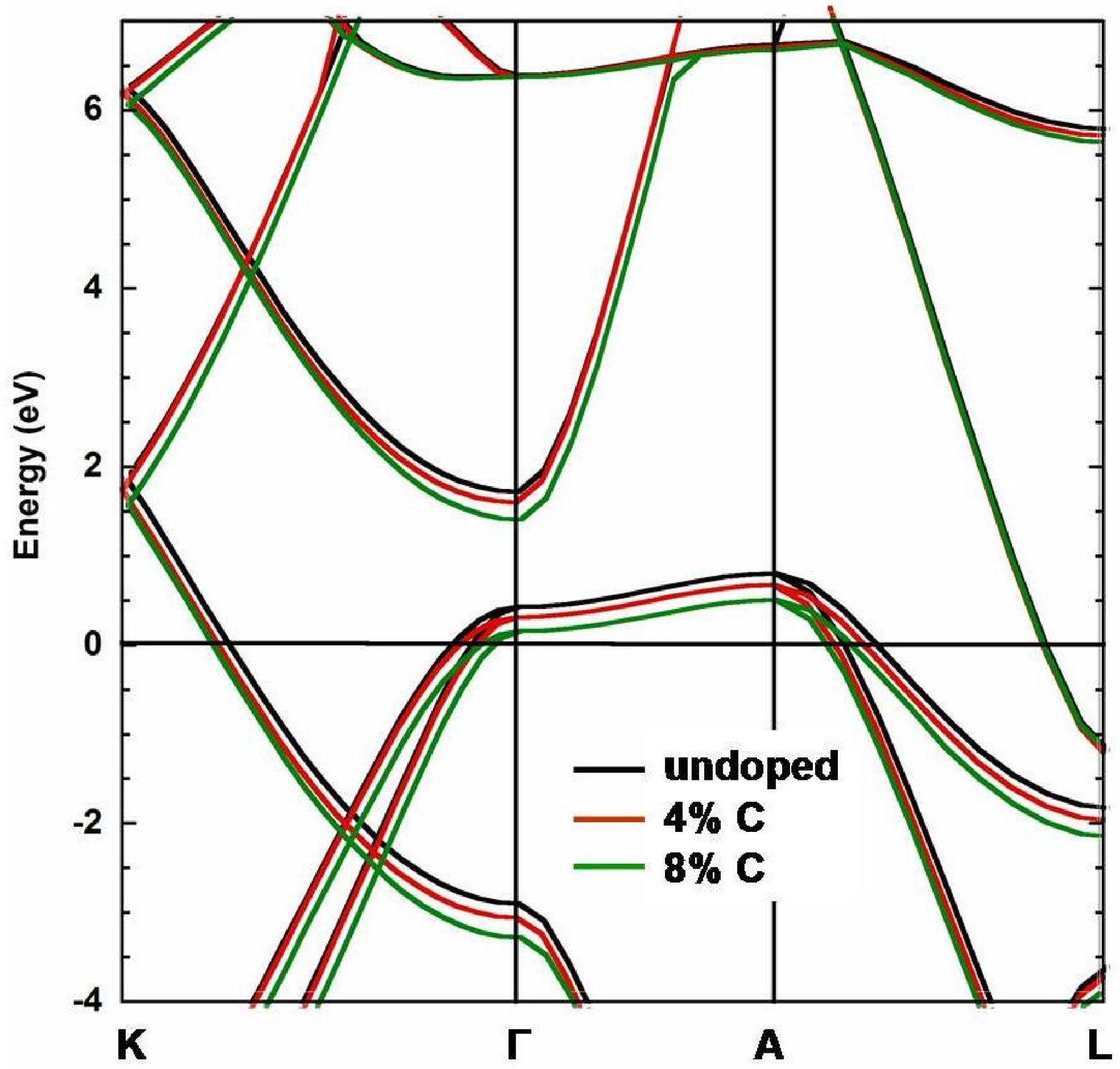}
\label{Fig:Bandstructure-C}
}
\caption{\textbf{Color:} Calculated a) partial density of states and b) bandstructure of pristine and C-doped \MB\ showing the \Sig\ and \PI-band shift for the different doping concentrations.}
\end{center}
\end{figure}

The DOS for the \PI\ and the \Sig-states of C-doped \MB\ are shown in Fig.~\ref{Fig:DOS-C}. Similarly to DOS of \MAB\, the high intensity in the \Sig-states close to the Fermi-level is decreasing significantly upon C-doping. While the hole-states at $E_F$ shift below the Fermi-level with increasing C-concentration, the \Ss-peak at $\sim$6 eV above $E_F$ does not seem to shift at all. However, the large peak in the \Sig-DOS at 17.5 eV shows several changes with increasing C concentration. The high intensity at 17.5 eV is decreasing in energy to 16.5 eV in \MAcB, while the less intense peak  18.2 eV in \MB\ increases significantly in intensity to form the largest peak in the \Sig-DOS in \MAcB. These changes can explain the apparent shift of peak \textit{c} Fig.~\ref{Fig:C001-exp}. The band structure for \MB, \MAbB, and \MAcB\ (shown in Fig.~\ref{Fig:Bandstructure-C}) further highlights the changes seen in the experiments and in the DOS. Firstly, the flat \Sig-band close to $E_F$ is slowly shifting below the Fermi-level with increasing C-concentration, resulting in  the \Sig-band nearly reaching $E_F$ at the $\Gamma$-point for \MAcB. However, the \Ss-band does not seem to shift at all in energy for higher C-doping. Interestingly, the \PI-band can be seen to shift downward by $0.25$ eV at $K$ for \MAcB. In contrast to the band-filling picture for Al-doping, the shifts of both the \Sig- and the \PI-bands close to $E_F$ in C-doped \MB\ indicate that the additional electrons provided by the C atoms replacing B fill both the \PI\ and the \Sig-states at the Fermi-level. This filling of both bands in the case of C-doping can explain why the relative pre-peak intensity of the B $K$ edge overestimates the amount of C needed for $T_c$ to reach 0 K.

\section*{Conclusions}

In conclusion, we have shown that the combination of high-resolution Z-contrast imaging and EELS can be used to study the effects of electron doping on the \Sig- and the \PI\ band of the 2-band superconductor \MB. We have further shown that substituting either C for B, or Al for Mg does not results in any long-range ordered superstructures visible by TEM, in contrast to previously reports for high concentration of Al-doping concentrations.\cite{Slusky01, Zandbergen02} Additionally, EELS-spectroscopy revealed that the effect of electron doping on the electronic structure of \MB\ is very similar for either dopant, while the superconducting transport properties of both sample are significantly different.
We have presented both theoretical and experimental evidence that doping \MB\ with electrons on the Mg and on the B-sites results filling the \Sig\ hole-states close to the Fermi-level, and thus a increase in the Fermi-level energy.
This filling of the superconducting hole-state in the \Sig\ band with increasing doping concentration has been shown to be closely linked to the disappearance of the \Sig\ band superconductivity once all these states are filled. Moreover, we have shown that the band-filling for Al and C-doping does not occur in a rigid band fashion; the \Ss-bands in Al-doped \MB\ shift faster than than \Sig-bands close to $E_F$, while in C-doped \MB\ the \Ss-bands remain constant and the \Sig-bands close the the Fermi-level shift significantly.

In addition, our results clearly show that selective doping on either the Mg or the B-site affects both bands of the B $p$-states. More specifically, it was previously  shown that there is a significant charge-transfer of electrons from the B-planes towards the Mg-planes\cite{WuMgB204} and adding extra electrons on either lattice site should result in filling of both the \Sig- and the \PI\ band. Interestingly, we could show that only C-doping results in a measurable filling of the \PI-band, while Al-doping does not influence the \PI-band DOS at all. Finally, we have presented experimental and theoretical evidence that the reduction of $T_c$ for both Al and C-doped \MB\ can be explained in large parts by the band-filling mechanism alone. Although an increase in inter-band scattering is not needed to explain the drop in $T_c$ of our \MABC\ samples, we can not affirmatively exclude effects such as increased inter-band scattering, local clustering of C atoms or the buckling of C-doped boron planes to occur in \MBC\ and cause a further decrease $T_c$.
However, in contrast to Al-doped \MB, the addition of C to \MB\ has been shown to increase the upper critical field (\HT) significantly, which can be explained by the increase electron-hole scattering in the B-planes and also results in a decrease in $T_c$. Our results presented in this paper suggest that adding additional hole on the Mg-sites by doping for example with Na to off-set the negative effects of C-doping on the \Sig-band, while keeping the C on the B-site to increase \HT\ but without the loss of $T_c$, thus resulting in a new high-field, high-$T_c$ superconductor

\section*{Acknowledgements} This manuscript has been authored by Brookhaven Science Associates, LLC under Contract No. DE-AC02-98CH1-886 and DOE FG02 96ER45610 with the U.S. Department of Energy, Division of Materials Sciences, Office of Basic Energy Science.

One author (RFK) acknowledges support as a Goldhaber Fellow at BNL. Other authors (LDC and AJZ) acknowledge additional support from the BNL-LDRD programs.


\begin{thebibliography}{43}
\expandafter\ifx\csname natexlab\endcsname\relax\def\natexlab#1{#1}\fi
\expandafter\ifx\csname bibnamefont\endcsname\relax
  \def\bibnamefont#1{#1}\fi
\expandafter\ifx\csname bibfnamefont\endcsname\relax
  \def\bibfnamefont#1{#1}\fi
\expandafter\ifx\csname citenamefont\endcsname\relax
  \def\citenamefont#1{#1}\fi
\expandafter\ifx\csname url\endcsname\relax
  \def\url#1{\texttt{#1}}\fi
\expandafter\ifx\csname urlprefix\endcsname\relax\def\urlprefix{URL }\fi
\providecommand{\bibinfo}[2]{#2}
\providecommand{\eprint}[2][]{\url{#2}}

\bibitem[{\citenamefont{Nagamatsu et~al.}(2001)\citenamefont{Nagamatsu,
  Nakagawa, Muranaka, Zenitani, and Akimitsu}}]{Nagamatsu01}
\bibinfo{author}{\bibfnamefont{J.}~\bibnamefont{Nagamatsu}},
  \bibinfo{author}{\bibfnamefont{N.}~\bibnamefont{Nakagawa}},
  \bibinfo{author}{\bibfnamefont{T.}~\bibnamefont{Muranaka}},
  \bibinfo{author}{\bibfnamefont{Y.}~\bibnamefont{Zenitani}}, \bibnamefont{and}
  \bibinfo{author}{\bibfnamefont{J.}~\bibnamefont{Akimitsu}},
  \bibinfo{journal}{Nature} \textbf{\bibinfo{volume}{410}}, \bibinfo{pages}{63}
  (\bibinfo{year}{2001}).

\bibitem[{\citenamefont{Brinkman et~al.}(2002)\citenamefont{Brinkman, Golubov,
  Rogalla, Dolgov, Kortus, Kong, Jepsen, and Andersen}}]{Brinkman02}
\bibinfo{author}{\bibfnamefont{A.}~\bibnamefont{Brinkman}},
  \bibinfo{author}{\bibfnamefont{A.~A.} \bibnamefont{Golubov}},
  \bibinfo{author}{\bibfnamefont{H.}~\bibnamefont{Rogalla}},
  \bibinfo{author}{\bibfnamefont{O.~V.} \bibnamefont{Dolgov}},
  \bibinfo{author}{\bibfnamefont{J.}~\bibnamefont{Kortus}},
  \bibinfo{author}{\bibfnamefont{Y.}~\bibnamefont{Kong}},
  \bibinfo{author}{\bibfnamefont{O.}~\bibnamefont{Jepsen}}, \bibnamefont{and}
  \bibinfo{author}{\bibfnamefont{O.~K.} \bibnamefont{Andersen}},
  \bibinfo{journal}{Physical Review B} \textbf{\bibinfo{volume}{65}},
  \bibinfo{pages}{180517(R)} (\bibinfo{year}{2002}).

\bibitem[{\citenamefont{Liu et~al.}(2001)\citenamefont{Liu, Mazin, and
  Kortus}}]{LiuPRL01}
\bibinfo{author}{\bibfnamefont{A.~Y.} \bibnamefont{Liu}},
  \bibinfo{author}{\bibfnamefont{I.~I.} \bibnamefont{Mazin}}, \bibnamefont{and}
  \bibinfo{author}{\bibfnamefont{J.}~\bibnamefont{Kortus}},
  \bibinfo{journal}{Physical Review Letters} \textbf{\bibinfo{volume}{87}},
  \bibinfo{pages}{087005} (\bibinfo{year}{2001}).

\bibitem[{\citenamefont{Agrestini et~al.}(2001)\citenamefont{Agrestini,
  Di~Castro, Sansone, Saini, Saccone, De~Negri, Giovannini, Colapietro, and
  Bianconi}}]{Agrestini01}
\bibinfo{author}{\bibfnamefont{S.}~\bibnamefont{Agrestini}},
  \bibinfo{author}{\bibfnamefont{D.}~\bibnamefont{Di~Castro}},
  \bibinfo{author}{\bibfnamefont{M.}~\bibnamefont{Sansone}},
  \bibinfo{author}{\bibfnamefont{N.~L.} \bibnamefont{Saini}},
  \bibinfo{author}{\bibfnamefont{A.}~\bibnamefont{Saccone}},
  \bibinfo{author}{\bibfnamefont{S.}~\bibnamefont{De~Negri}},
  \bibinfo{author}{\bibfnamefont{M.}~\bibnamefont{Giovannini}},
  \bibinfo{author}{\bibfnamefont{M.}~\bibnamefont{Colapietro}},
  \bibnamefont{and} \bibinfo{author}{\bibfnamefont{A.}~\bibnamefont{Bianconi}},
  \bibinfo{journal}{Journal Of Physics-Condensed Matter}
  \textbf{\bibinfo{volume}{13}}, \bibinfo{pages}{11689} (\bibinfo{year}{2001}).

\bibitem[{\citenamefont{Papagelis et~al.}(2003)\citenamefont{Papagelis,
  Arvanitidis, Prassides, Schenck, Takenobu, and Iwasa}}]{Papagelis03}
\bibinfo{author}{\bibfnamefont{K.}~\bibnamefont{Papagelis}},
  \bibinfo{author}{\bibfnamefont{J.}~\bibnamefont{Arvanitidis}},
  \bibinfo{author}{\bibfnamefont{K.}~\bibnamefont{Prassides}},
  \bibinfo{author}{\bibfnamefont{A.}~\bibnamefont{Schenck}},
  \bibinfo{author}{\bibfnamefont{T.}~\bibnamefont{Takenobu}}, \bibnamefont{and}
  \bibinfo{author}{\bibfnamefont{Y.}~\bibnamefont{Iwasa}},
  \bibinfo{journal}{Europhysics Letters} \textbf{\bibinfo{volume}{61}},
  \bibinfo{pages}{254} (\bibinfo{year}{2003}).

\bibitem[{\citenamefont{Schmidt et~al.}(2003)\citenamefont{Schmidt,
  Zasadzinski, Gray, and Hinks}}]{Schmidt03}
\bibinfo{author}{\bibfnamefont{H.}~\bibnamefont{Schmidt}},
  \bibinfo{author}{\bibfnamefont{J.~F.} \bibnamefont{Zasadzinski}},
  \bibinfo{author}{\bibfnamefont{K.~E.} \bibnamefont{Gray}}, \bibnamefont{and}
  \bibinfo{author}{\bibfnamefont{D.~G.} \bibnamefont{Hinks}},
  \bibinfo{journal}{Physica C: Superconductivity}
  \textbf{\bibinfo{volume}{385}}, \bibinfo{pages}{221} (\bibinfo{year}{2003}).

\bibitem[{\citenamefont{Di~Castro et~al.}(2002)\citenamefont{Di~Castro,
  Agrestini, Campi, Cassetta, Colapietro, Congeduti, Continenza, De~Negri,
  Giovannini, Massidda et~al.}}]{DiCastro02}
\bibinfo{author}{\bibfnamefont{D.}~\bibnamefont{Di~Castro}},
  \bibinfo{author}{\bibfnamefont{S.}~\bibnamefont{Agrestini}},
  \bibinfo{author}{\bibfnamefont{G.}~\bibnamefont{Campi}},
  \bibinfo{author}{\bibfnamefont{A.}~\bibnamefont{Cassetta}},
  \bibinfo{author}{\bibfnamefont{M.}~\bibnamefont{Colapietro}},
  \bibinfo{author}{\bibfnamefont{A.}~\bibnamefont{Congeduti}},
  \bibinfo{author}{\bibfnamefont{A.}~\bibnamefont{Continenza}},
  \bibinfo{author}{\bibfnamefont{S.}~\bibnamefont{De~Negri}},
  \bibinfo{author}{\bibfnamefont{M.}~\bibnamefont{Giovannini}},
  \bibinfo{author}{\bibfnamefont{S.}~\bibnamefont{Massidda}},
  \bibnamefont{et~al.}, \bibinfo{journal}{Europhysics Letters}
  \textbf{\bibinfo{volume}{58}}, \bibinfo{pages}{278} (\bibinfo{year}{2002}).

\bibitem[{\citenamefont{Postorino et~al.}(2001)\citenamefont{Postorino,
  Congeduti, Dore, Nucara, Bianconi, Di~Castro, De~Negri, and
  Saccone}}]{Postorino02}
\bibinfo{author}{\bibfnamefont{P.}~\bibnamefont{Postorino}},
  \bibinfo{author}{\bibfnamefont{A.}~\bibnamefont{Congeduti}},
  \bibinfo{author}{\bibfnamefont{P.}~\bibnamefont{Dore}},
  \bibinfo{author}{\bibfnamefont{A.}~\bibnamefont{Nucara}},
  \bibinfo{author}{\bibfnamefont{A.}~\bibnamefont{Bianconi}},
  \bibinfo{author}{\bibfnamefont{D.}~\bibnamefont{Di~Castro}},
  \bibinfo{author}{\bibfnamefont{S.}~\bibnamefont{De~Negri}}, \bibnamefont{and}
  \bibinfo{author}{\bibfnamefont{A.}~\bibnamefont{Saccone}},
  \bibinfo{journal}{Physical Review B} \textbf{\bibinfo{volume}{65}},
  \bibinfo{pages}{020507(R)} (\bibinfo{year}{2001}).

\bibitem[{\citenamefont{Bianconi et~al.}(2002)\citenamefont{Bianconi,
  Agrestini, Di~Castro, Campi, Zangari, Saini, Saccone, De~Negri, Giovannini,
  Profeta et~al.}}]{Bianconi02}
\bibinfo{author}{\bibfnamefont{A.}~\bibnamefont{Bianconi}},
  \bibinfo{author}{\bibfnamefont{S.}~\bibnamefont{Agrestini}},
  \bibinfo{author}{\bibfnamefont{D.}~\bibnamefont{Di~Castro}},
  \bibinfo{author}{\bibfnamefont{G.}~\bibnamefont{Campi}},
  \bibinfo{author}{\bibfnamefont{G.}~\bibnamefont{Zangari}},
  \bibinfo{author}{\bibfnamefont{N.~L.} \bibnamefont{Saini}},
  \bibinfo{author}{\bibfnamefont{A.}~\bibnamefont{Saccone}},
  \bibinfo{author}{\bibfnamefont{S.}~\bibnamefont{De~Negri}},
  \bibinfo{author}{\bibfnamefont{M.}~\bibnamefont{Giovannini}},
  \bibinfo{author}{\bibfnamefont{G.}~\bibnamefont{Profeta}},
  \bibnamefont{et~al.}, \bibinfo{journal}{Physical Review B}
  \textbf{\bibinfo{volume}{65}}, \bibinfo{pages}{174515}
  (\bibinfo{year}{2002}).

\bibitem[{\citenamefont{Zambano and Cooley}(2005)}]{Zambano05}
\bibinfo{author}{\bibfnamefont{A.}~\bibnamefont{Zambano}} \bibnamefont{and}
  \bibinfo{author}{\bibfnamefont{L.~D.} \bibnamefont{Cooley}}
  (\bibinfo{year}{2005}).

\bibitem[{\citenamefont{Papavassiliou et~al.}(2002)\citenamefont{Papavassiliou,
  Pissas, Karayanni, Fardis, Koutandos, and Prassides}}]{Papavassiliou02}
\bibinfo{author}{\bibfnamefont{G.}~\bibnamefont{Papavassiliou}},
  \bibinfo{author}{\bibfnamefont{M.}~\bibnamefont{Pissas}},
  \bibinfo{author}{\bibfnamefont{M.}~\bibnamefont{Karayanni}},
  \bibinfo{author}{\bibfnamefont{M.}~\bibnamefont{Fardis}},
  \bibinfo{author}{\bibfnamefont{S.}~\bibnamefont{Koutandos}},
  \bibnamefont{and}
  \bibinfo{author}{\bibfnamefont{K.}~\bibnamefont{Prassides}},
  \bibinfo{journal}{Physical Review B} \textbf{\bibinfo{volume}{66}},
  \bibinfo{pages}{140514(R)} (\bibinfo{year}{2002}).

\bibitem[{\citenamefont{Margadonna et~al.}(2002)\citenamefont{Margadonna,
  Prassides, Arvanitidis, Pissas, Papavassiliou, and Fitch}}]{Margadonna02}
\bibinfo{author}{\bibfnamefont{S.}~\bibnamefont{Margadonna}},
  \bibinfo{author}{\bibfnamefont{K.}~\bibnamefont{Prassides}},
  \bibinfo{author}{\bibfnamefont{I.}~\bibnamefont{Arvanitidis}},
  \bibinfo{author}{\bibfnamefont{M.}~\bibnamefont{Pissas}},
  \bibinfo{author}{\bibfnamefont{G.}~\bibnamefont{Papavassiliou}},
  \bibnamefont{and} \bibinfo{author}{\bibfnamefont{A.~N.} \bibnamefont{Fitch}},
  \bibinfo{journal}{Physical Review B} \textbf{\bibinfo{volume}{66}},
  \bibinfo{pages}{014518} (\bibinfo{year}{2002}).

\bibitem[{\citenamefont{Putti et~al.}(2003)\citenamefont{Putti, Affronte,
  Manfrinetti, and Palenzona}}]{Putti03}
\bibinfo{author}{\bibfnamefont{M.}~\bibnamefont{Putti}},
  \bibinfo{author}{\bibfnamefont{M.}~\bibnamefont{Affronte}},
  \bibinfo{author}{\bibfnamefont{P.}~\bibnamefont{Manfrinetti}},
  \bibnamefont{and}
  \bibinfo{author}{\bibfnamefont{A.}~\bibnamefont{Palenzona}},
  \bibinfo{journal}{Physical Review B} \textbf{\bibinfo{volume}{68}},
  \bibinfo{pages}{094514} (\bibinfo{year}{2003}).

\bibitem[{\citenamefont{Putti et~al.}(2005)\citenamefont{Putti, Ferdeghini,
  Monni, Pallecchi, Tarantini, Manfrinetti, Palenzona, Daghero, Gonnelli, and
  Stepanov}}]{Putti05}
\bibinfo{author}{\bibfnamefont{M.}~\bibnamefont{Putti}},
  \bibinfo{author}{\bibfnamefont{C.}~\bibnamefont{Ferdeghini}},
  \bibinfo{author}{\bibfnamefont{M.}~\bibnamefont{Monni}},
  \bibinfo{author}{\bibfnamefont{I.}~\bibnamefont{Pallecchi}},
  \bibinfo{author}{\bibfnamefont{C.}~\bibnamefont{Tarantini}},
  \bibinfo{author}{\bibfnamefont{P.}~\bibnamefont{Manfrinetti}},
  \bibinfo{author}{\bibfnamefont{A.}~\bibnamefont{Palenzona}},
  \bibinfo{author}{\bibfnamefont{D.}~\bibnamefont{Daghero}},
  \bibinfo{author}{\bibfnamefont{R.~S.} \bibnamefont{Gonnelli}},
  \bibnamefont{and} \bibinfo{author}{\bibfnamefont{V.~A.}
  \bibnamefont{Stepanov}}, \bibinfo{journal}{Physical Review B}
  \textbf{\bibinfo{volume}{71}}, \bibinfo{pages}{144505}
  (\bibinfo{year}{2005}).

\bibitem[{\citenamefont{Kortus et~al.}(2005)\citenamefont{Kortus, Dolgov,
  Kremer, and Golubov}}]{Kortus05}
\bibinfo{author}{\bibfnamefont{J.}~\bibnamefont{Kortus}},
  \bibinfo{author}{\bibfnamefont{O.~V.} \bibnamefont{Dolgov}},
  \bibinfo{author}{\bibfnamefont{R.~K.} \bibnamefont{Kremer}},
  \bibnamefont{and} \bibinfo{author}{\bibfnamefont{A.~A.}
  \bibnamefont{Golubov}}, \bibinfo{journal}{Physical Review Letters}
  \textbf{\bibinfo{volume}{94}}, \bibinfo{pages}{027002}
  (\bibinfo{year}{2005}).

\bibitem[{\citenamefont{Gonnelli et~al.}(2005)\citenamefont{Gonnelli, Daghero,
  Calzolari, Ummarino, Dellarocca, Stepanov, Kazakov, Zhigadlo, and
  Karpinski}}]{Gonnelli05}
\bibinfo{author}{\bibfnamefont{R.~S.} \bibnamefont{Gonnelli}},
  \bibinfo{author}{\bibfnamefont{D.}~\bibnamefont{Daghero}},
  \bibinfo{author}{\bibfnamefont{A.}~\bibnamefont{Calzolari}},
  \bibinfo{author}{\bibfnamefont{G.~A.} \bibnamefont{Ummarino}},
  \bibinfo{author}{\bibfnamefont{V.}~\bibnamefont{Dellarocca}},
  \bibinfo{author}{\bibfnamefont{V.~A.} \bibnamefont{Stepanov}},
  \bibinfo{author}{\bibfnamefont{S.~M.} \bibnamefont{Kazakov}},
  \bibinfo{author}{\bibfnamefont{N.}~\bibnamefont{Zhigadlo}}, \bibnamefont{and}
  \bibinfo{author}{\bibfnamefont{J.}~\bibnamefont{Karpinski}},
  \bibinfo{journal}{Physical Review B} \textbf{\bibinfo{volume}{71}},
  \bibinfo{pages}{060503(R)} (\bibinfo{year}{2005}).

\bibitem[{\citenamefont{Masui et~al.}(2004)\citenamefont{Masui, Lee, and
  Tajima}}]{Matsui04}
\bibinfo{author}{\bibfnamefont{T.}~\bibnamefont{Masui}},
  \bibinfo{author}{\bibfnamefont{S.}~\bibnamefont{Lee}}, \bibnamefont{and}
  \bibinfo{author}{\bibfnamefont{S.}~\bibnamefont{Tajima}},
  \bibinfo{journal}{Phys.~Rev.~B} \textbf{\bibinfo{volume}{70}},
  \bibinfo{pages}{024504} (\bibinfo{year}{2004}).

\bibitem[{\citenamefont{Tomita et~al.}(2001)\citenamefont{Tomita, Hamlin,
  Schilling, Hinks, and Jorgensen}}]{Tomita01}
\bibinfo{author}{\bibfnamefont{T.}~\bibnamefont{Tomita}},
  \bibinfo{author}{\bibfnamefont{J.~J.} \bibnamefont{Hamlin}},
  \bibinfo{author}{\bibfnamefont{J.~S.} \bibnamefont{Schilling}},
  \bibinfo{author}{\bibfnamefont{D.~G.} \bibnamefont{Hinks}}, \bibnamefont{and}
  \bibinfo{author}{\bibfnamefont{J.~D.} \bibnamefont{Jorgensen}},
  \bibinfo{journal}{Phys.~Rev.~B} \textbf{\bibinfo{volume}{64}},
  \bibinfo{pages}{092505} (\bibinfo{year}{2001}).

\bibitem[{\citenamefont{Pogrebnyakov et~al.}(2004)\citenamefont{Pogrebnyakov,
  Xi, Redwing, Vaithyanathan, Schlom, Soukiassian, Mi, Jia, Giencke, Eom
  et~al.}}]{Pogrebnyakov04}
\bibinfo{author}{\bibfnamefont{A.~V.} \bibnamefont{Pogrebnyakov}},
  \bibinfo{author}{\bibfnamefont{X.~X.} \bibnamefont{Xi}},
  \bibinfo{author}{\bibfnamefont{J.~M.} \bibnamefont{Redwing}},
  \bibinfo{author}{\bibfnamefont{V.}~\bibnamefont{Vaithyanathan}},
  \bibinfo{author}{\bibfnamefont{D.~G.} \bibnamefont{Schlom}},
  \bibinfo{author}{\bibfnamefont{A.}~\bibnamefont{Soukiassian}},
  \bibinfo{author}{\bibfnamefont{S.~B.} \bibnamefont{Mi}},
  \bibinfo{author}{\bibfnamefont{C.~L.} \bibnamefont{Jia}},
  \bibinfo{author}{\bibfnamefont{J.~E.} \bibnamefont{Giencke}},
  \bibinfo{author}{\bibfnamefont{C.~B.} \bibnamefont{Eom}},
  \bibnamefont{et~al.}, \bibinfo{journal}{Appl.~Phys.~Lett.}
  \textbf{\bibinfo{volume}{85}}, \bibinfo{pages}{2017} (\bibinfo{year}{2004}).

\bibitem[{\citenamefont{Karpinski et~al.}(2005)\citenamefont{Karpinski,
  Zhugadlo, Schuck, Kazakov, Batlogg, Rogacki, Puzniak, Jun, M\"{u}ller,
  W\"{a}gli et~al.}}]{Karpinski05}
\bibinfo{author}{\bibfnamefont{J.}~\bibnamefont{Karpinski}},
  \bibinfo{author}{\bibfnamefont{N.~D.} \bibnamefont{Zhugadlo}},
  \bibinfo{author}{\bibfnamefont{G.}~\bibnamefont{Schuck}},
  \bibinfo{author}{\bibfnamefont{S.~M.} \bibnamefont{Kazakov}},
  \bibinfo{author}{\bibfnamefont{B.}~\bibnamefont{Batlogg}},
  \bibinfo{author}{\bibfnamefont{K.}~\bibnamefont{Rogacki}},
  \bibinfo{author}{\bibfnamefont{R.}~\bibnamefont{Puzniak}},
  \bibinfo{author}{\bibfnamefont{J.}~\bibnamefont{Jun}},
  \bibinfo{author}{\bibfnamefont{E.}~\bibnamefont{M\"{u}ller}},
  \bibinfo{author}{\bibfnamefont{P.}~\bibnamefont{W\"{a}gli}},
  \bibnamefont{et~al.}, \bibinfo{journal}{Phys.\ Rev.\ B}
  \textbf{\bibinfo{volume}{71}}, \bibinfo{pages}{174506}
  (\bibinfo{year}{2005}).

\bibitem[{\citenamefont{Klie et~al.}(2003{\natexlab{a}})\citenamefont{Klie, Su,
  Zhu, Davenport, Idrobo, Browning, and Nellist}}]{klie03}
\bibinfo{author}{\bibfnamefont{R.~F.} \bibnamefont{Klie}},
  \bibinfo{author}{\bibfnamefont{H.}~\bibnamefont{Su}},
  \bibinfo{author}{\bibfnamefont{Y.}~\bibnamefont{Zhu}},
  \bibinfo{author}{\bibfnamefont{J.~W.} \bibnamefont{Davenport}},
  \bibinfo{author}{\bibfnamefont{J.~C.} \bibnamefont{Idrobo}},
  \bibinfo{author}{\bibfnamefont{N.~D.} \bibnamefont{Browning}},
  \bibnamefont{and} \bibinfo{author}{\bibfnamefont{P.~D.}
  \bibnamefont{Nellist}}, \bibinfo{journal}{Phys.\ Rev.\ B}
  \textbf{\bibinfo{volume}{67}}, \bibinfo{pages}{144508}
  (\bibinfo{year}{2003}{\natexlab{a}}).

\bibitem[{\citenamefont{Klie et~al.}(2002)\citenamefont{Klie, Idrobo, Browning,
  Serquis, Zhu, Liao, and Muller}}]{klie02APL}
\bibinfo{author}{\bibfnamefont{R.~F.} \bibnamefont{Klie}},
  \bibinfo{author}{\bibfnamefont{J.~C.} \bibnamefont{Idrobo}},
  \bibinfo{author}{\bibfnamefont{N.~D.} \bibnamefont{Browning}},
  \bibinfo{author}{\bibfnamefont{A.}~\bibnamefont{Serquis}},
  \bibinfo{author}{\bibfnamefont{Y.~T.} \bibnamefont{Zhu}},
  \bibinfo{author}{\bibfnamefont{X.~Z.} \bibnamefont{Liao}}, \bibnamefont{and}
  \bibinfo{author}{\bibfnamefont{F.~M.} \bibnamefont{Muller}},
  \bibinfo{journal}{Appl.\ Phys.\ Lett.} \textbf{\bibinfo{volume}{80}},
  \bibinfo{pages}{3970} (\bibinfo{year}{2002}).

\bibitem[{\citenamefont{Zhu et~al.}(2002)\citenamefont{Zhu, Moodenbaugh,
  Schneider, Davenport, Vogt, Li, Gu, Fischer, and Tafto}}]{zhu02}
\bibinfo{author}{\bibfnamefont{Y.}~\bibnamefont{Zhu}},
  \bibinfo{author}{\bibfnamefont{A.~R.} \bibnamefont{Moodenbaugh}},
  \bibinfo{author}{\bibfnamefont{G.}~\bibnamefont{Schneider}},
  \bibinfo{author}{\bibfnamefont{J.~W.} \bibnamefont{Davenport}},
  \bibinfo{author}{\bibfnamefont{T.}~\bibnamefont{Vogt}},
  \bibinfo{author}{\bibfnamefont{Q.}~\bibnamefont{Li}},
  \bibinfo{author}{\bibfnamefont{G.}~\bibnamefont{Gu}},
  \bibinfo{author}{\bibfnamefont{D.~A.} \bibnamefont{Fischer}},
  \bibnamefont{and} \bibinfo{author}{\bibfnamefont{J.}~\bibnamefont{Tafto}},
  \bibinfo{journal}{Phys.\ Rev.\ Lett.} \textbf{\bibinfo{volume}{88}},
  \bibinfo{pages}{247002} (\bibinfo{year}{2002}).

\bibitem[{\citenamefont{Cooley et~al.}(2005)\citenamefont{Cooley, Zambano,
  Moodenbaugh, Klie, Zheng, and Zhu}}]{Cooley05PRL}
\bibinfo{author}{\bibfnamefont{L.~D.} \bibnamefont{Cooley}},
  \bibinfo{author}{\bibfnamefont{A.~J.} \bibnamefont{Zambano}},
  \bibinfo{author}{\bibfnamefont{A.~R.} \bibnamefont{Moodenbaugh}},
  \bibinfo{author}{\bibfnamefont{R.~F.} \bibnamefont{Klie}},
  \bibinfo{author}{\bibfnamefont{J.-C.} \bibnamefont{Zheng}}, \bibnamefont{and}
  \bibinfo{author}{\bibfnamefont{Y.}~\bibnamefont{Zhu}} (\bibinfo{year}{2005}).

\bibitem[{\citenamefont{James and Browning}(1999)}]{james99-UM}
\bibinfo{author}{\bibfnamefont{E.~M.} \bibnamefont{James}} \bibnamefont{and}
  \bibinfo{author}{\bibfnamefont{N.~D.} \bibnamefont{Browning}},
  \bibinfo{journal}{Ultramicroscopy} \textbf{\bibinfo{volume}{78}},
  \bibinfo{pages}{125} (\bibinfo{year}{1999}).

\bibitem[{\citenamefont{James et~al.}(1998)\citenamefont{James, Browning,
  Nicholls, Kawasaki, Xin, and Stemmer}}]{james98-EM}
\bibinfo{author}{\bibfnamefont{E.~M.} \bibnamefont{James}},
  \bibinfo{author}{\bibfnamefont{N.~D.} \bibnamefont{Browning}},
  \bibinfo{author}{\bibfnamefont{A.~W.} \bibnamefont{Nicholls}},
  \bibinfo{author}{\bibfnamefont{M.}~\bibnamefont{Kawasaki}},
  \bibinfo{author}{\bibfnamefont{Y.}~\bibnamefont{Xin}}, \bibnamefont{and}
  \bibinfo{author}{\bibfnamefont{S.}~\bibnamefont{Stemmer}},
  \bibinfo{journal}{J.\ Electron Microsc.} \textbf{\bibinfo{volume}{47}},
  \bibinfo{pages}{561} (\bibinfo{year}{1998}).

\bibitem[{\citenamefont{Nellist and Pennycook}(1999)}]{nellist99}
\bibinfo{author}{\bibfnamefont{P.~D.} \bibnamefont{Nellist}} \bibnamefont{and}
  \bibinfo{author}{\bibfnamefont{S.~J.} \bibnamefont{Pennycook}},
  \bibinfo{journal}{Ultramicroscopy} \textbf{\bibinfo{volume}{78}},
  \bibinfo{pages}{111} (\bibinfo{year}{1999}).

\bibitem[{\citenamefont{Klie and Zhu}(2005)}]{Klie05Micron}
\bibinfo{author}{\bibfnamefont{R.~F.} \bibnamefont{Klie}} \bibnamefont{and}
  \bibinfo{author}{\bibfnamefont{Y.}~\bibnamefont{Zhu}},
  \bibinfo{journal}{Micron} \textbf{\bibinfo{volume}{36}}, \bibinfo{pages}{219}
  (\bibinfo{year}{2005}).

\bibitem[{\citenamefont{Cowley}(1986)}]{cowley86}
\bibinfo{author}{\bibfnamefont{J.~M.} \bibnamefont{Cowley}},
  \bibinfo{journal}{J.\ Electron.\ Microsc.\ Tech.}
  \textbf{\bibinfo{volume}{3}}, \bibinfo{pages}{25} (\bibinfo{year}{1986}).

\bibitem[{\citenamefont{Egerton}(1986)}]{egerton86}
\bibinfo{author}{\bibfnamefont{R.~F.} \bibnamefont{Egerton}},
  \emph{\bibinfo{title}{Electron Energy Loss Spectroscopy in the Electron
  Microscope}} (\bibinfo{publisher}{Plenum Press, New York},
  \bibinfo{year}{1986}).

\bibitem[{\citenamefont{Fertig and Rose}(1981)}]{fertig81}
\bibinfo{author}{\bibfnamefont{J.}~\bibnamefont{Fertig}} \bibnamefont{and}
  \bibinfo{author}{\bibfnamefont{H.}~\bibnamefont{Rose}},
  \bibinfo{journal}{Optik} \textbf{\bibinfo{volume}{59}}, \bibinfo{pages}{407}
  (\bibinfo{year}{1981}).

\bibitem[{\citenamefont{Browning et~al.}(1993)\citenamefont{Browning, Yuan, and
  Brown}}]{browning93}
\bibinfo{author}{\bibfnamefont{N.~D.} \bibnamefont{Browning}},
  \bibinfo{author}{\bibfnamefont{J.}~\bibnamefont{Yuan}}, \bibnamefont{and}
  \bibinfo{author}{\bibfnamefont{L.~M.} \bibnamefont{Brown}},
  \bibinfo{journal}{Phil.\ Mag.\ A} \textbf{\bibinfo{volume}{67}},
  \bibinfo{pages}{261} (\bibinfo{year}{1993}).

\bibitem[{\citenamefont{Hohenberg and Kohn}(1965)}]{Hohenberg65}
\bibinfo{author}{\bibfnamefont{P.}~\bibnamefont{Hohenberg}} \bibnamefont{and}
  \bibinfo{author}{\bibfnamefont{W.}~\bibnamefont{Kohn}},
  \bibinfo{journal}{Phys.~Rev.} \textbf{\bibinfo{volume}{136}},
  \bibinfo{pages}{B864} (\bibinfo{year}{1965}).

\bibitem[{\citenamefont{Parr and Yang}(1989)}]{Parr89}
\bibinfo{author}{\bibfnamefont{G.~G.} \bibnamefont{Parr}} \bibnamefont{and}
  \bibinfo{author}{\bibfnamefont{W.~T.} \bibnamefont{Yang}},
  \emph{\bibinfo{title}{Density-functional theory of atoms and molecules}}
  (\bibinfo{publisher}{Oxford University Press}, \bibinfo{year}{1989}).

\bibitem[{\citenamefont{Blaha et~al.}(2001)\citenamefont{Blaha, Schwarz,
  Madsen, Kvasnicka, and Luitz}}]{Blaha01}
\bibinfo{author}{\bibfnamefont{P.}~\bibnamefont{Blaha}},
  \bibinfo{author}{\bibfnamefont{K.}~\bibnamefont{Schwarz}},
  \bibinfo{author}{\bibfnamefont{G.}~\bibnamefont{Madsen}},
  \bibinfo{author}{\bibfnamefont{D.}~\bibnamefont{Kvasnicka}},
  \bibnamefont{and} \bibinfo{author}{\bibfnamefont{J.}~\bibnamefont{Luitz}},
  \emph{\bibinfo{title}{WIEN2k, An Augmented Plane Wave + Local Orbitals
  Program for Calculating Crystal Properties}} (\bibinfo{publisher}{Techn.
  Universitat Wien}, \bibinfo{year}{2001}).

\bibitem[{\citenamefont{Perdew et~al.}(1996)\citenamefont{Perdew, Burke, and
  Ernzerhof}}]{Perdew96}
\bibinfo{author}{\bibfnamefont{J.~P.} \bibnamefont{Perdew}},
  \bibinfo{author}{\bibfnamefont{K.}~\bibnamefont{Burke}}, \bibnamefont{and}
  \bibinfo{author}{\bibfnamefont{M.}~\bibnamefont{Ernzerhof}},
  \bibinfo{journal}{Phys.~Rev.~Lett.} \textbf{\bibinfo{volume}{77}},
  \bibinfo{pages}{3865} (\bibinfo{year}{1996}).

\bibitem[{\citenamefont{Wilke et~al.}(2004)\citenamefont{Wilke, Bud'ko,
  Canfield, Finnemore, Suplinskas, and Hannahs}}]{Wilke04}
\bibinfo{author}{\bibfnamefont{R.~H.~T.} \bibnamefont{Wilke}},
  \bibinfo{author}{\bibfnamefont{S.~L.} \bibnamefont{Bud'ko}},
  \bibinfo{author}{\bibfnamefont{P.~C.} \bibnamefont{Canfield}},
  \bibinfo{author}{\bibfnamefont{D.~K.} \bibnamefont{Finnemore}},
  \bibinfo{author}{\bibfnamefont{R.~J.} \bibnamefont{Suplinskas}},
  \bibnamefont{and} \bibinfo{author}{\bibfnamefont{S.~T.}
  \bibnamefont{Hannahs}}, \bibinfo{journal}{Physical Review Letters}
  \textbf{\bibinfo{volume}{92}}, \bibinfo{pages}{217003}
  (\bibinfo{year}{2004}).

\bibitem[{\citenamefont{Klie et~al.}(2003{\natexlab{b}})\citenamefont{Klie,
  Zhu, Schneider, and Tafto}}]{klie03-MgB2}
\bibinfo{author}{\bibfnamefont{R.~F.} \bibnamefont{Klie}},
  \bibinfo{author}{\bibfnamefont{Y.}~\bibnamefont{Zhu}},
  \bibinfo{author}{\bibfnamefont{G.}~\bibnamefont{Schneider}},
  \bibnamefont{and} \bibinfo{author}{\bibfnamefont{J.}~\bibnamefont{Tafto}},
  \bibinfo{journal}{Appl.\ Phys.\ Lett.} \textbf{\bibinfo{volume}{82}},
  \bibinfo{pages}{4316} (\bibinfo{year}{2003}{\natexlab{b}}).

\bibitem[{\citenamefont{Idrobo et~al.}(2004)\citenamefont{Idrobo, Ogut,
  Yildirim, Klie, and Browning}}]{Idrobo05}
\bibinfo{author}{\bibfnamefont{J.~C.} \bibnamefont{Idrobo}},
  \bibinfo{author}{\bibfnamefont{S.}~\bibnamefont{Ogut}},
  \bibinfo{author}{\bibfnamefont{T.}~\bibnamefont{Yildirim}},
  \bibinfo{author}{\bibfnamefont{R.~F.} \bibnamefont{Klie}}, \bibnamefont{and}
  \bibinfo{author}{\bibfnamefont{N.~D.} \bibnamefont{Browning}},
  \bibinfo{journal}{Physical Review B} \textbf{\bibinfo{volume}{70}},
  \bibinfo{pages}{172503} (\bibinfo{year}{2004}).

\bibitem[{\citenamefont{Brydson et~al.}(1989)\citenamefont{Brydson, Sauer,
  Engel, Thomas, Zeitler, Kosugi, and Kuroda}}]{Brydson89}
\bibinfo{author}{\bibfnamefont{R.}~\bibnamefont{Brydson}},
  \bibinfo{author}{\bibfnamefont{H.}~\bibnamefont{Sauer}},
  \bibinfo{author}{\bibfnamefont{W.}~\bibnamefont{Engel}},
  \bibinfo{author}{\bibfnamefont{J.~M.} \bibnamefont{Thomas}},
  \bibinfo{author}{\bibfnamefont{E.}~\bibnamefont{Zeitler}},
  \bibinfo{author}{\bibfnamefont{N.}~\bibnamefont{Kosugi}}, \bibnamefont{and}
  \bibinfo{author}{\bibfnamefont{H.}~\bibnamefont{Kuroda}},
  \bibinfo{journal}{J.~Phys.: Condens.~Matte} \textbf{\bibinfo{volume}{1}},
  \bibinfo{pages}{797} (\bibinfo{year}{1989}).

\bibitem[{\citenamefont{Slusky et~al.}(2001)\citenamefont{Slusky, Rogado,
  Regan, Hayward, Khalifah, He, Inumaru, Loureiro, Haas, Zandbergen
  et~al.}}]{Slusky01}
\bibinfo{author}{\bibfnamefont{J.~S.} \bibnamefont{Slusky}},
  \bibinfo{author}{\bibfnamefont{N.}~\bibnamefont{Rogado}},
  \bibinfo{author}{\bibfnamefont{K.~A.} \bibnamefont{Regan}},
  \bibinfo{author}{\bibfnamefont{M.~A.} \bibnamefont{Hayward}},
  \bibinfo{author}{\bibfnamefont{P.}~\bibnamefont{Khalifah}},
  \bibinfo{author}{\bibfnamefont{T.}~\bibnamefont{He}},
  \bibinfo{author}{\bibfnamefont{K.}~\bibnamefont{Inumaru}},
  \bibinfo{author}{\bibfnamefont{S.~M.} \bibnamefont{Loureiro}},
  \bibinfo{author}{\bibfnamefont{M.~K.} \bibnamefont{Haas}},
  \bibinfo{author}{\bibfnamefont{H.~W.} \bibnamefont{Zandbergen}},
  \bibnamefont{et~al.}, \bibinfo{journal}{Nature}
  \textbf{\bibinfo{volume}{410}}, \bibinfo{pages}{343} (\bibinfo{year}{2001}).

\bibitem[{\citenamefont{Zandbergen et~al.}(2002)\citenamefont{Zandbergen, Wu,
  Jiang, Hayward, Haas, and Cava}}]{Zandbergen02}
\bibinfo{author}{\bibfnamefont{H.~W.} \bibnamefont{Zandbergen}},
  \bibinfo{author}{\bibfnamefont{M.~Y.} \bibnamefont{Wu}},
  \bibinfo{author}{\bibfnamefont{H.}~\bibnamefont{Jiang}},
  \bibinfo{author}{\bibfnamefont{M.~A.} \bibnamefont{Hayward}},
  \bibinfo{author}{\bibfnamefont{M.~K.} \bibnamefont{Haas}}, \bibnamefont{and}
  \bibinfo{author}{\bibfnamefont{R.~J.} \bibnamefont{Cava}},
  \bibinfo{journal}{Physica C-Superconductivity And Its Applications}
  \textbf{\bibinfo{volume}{366}}, \bibinfo{pages}{221} (\bibinfo{year}{2002}).

\bibitem[{\citenamefont{Wu et~al.}(2004)\citenamefont{Wu, Zhu, Vogt, Su,
  Davenport, and Tafto}}]{WuMgB204}
\bibinfo{author}{\bibfnamefont{L.~J.} \bibnamefont{Wu}},
  \bibinfo{author}{\bibfnamefont{Y.~M.} \bibnamefont{Zhu}},
  \bibinfo{author}{\bibfnamefont{T.}~\bibnamefont{Vogt}},
  \bibinfo{author}{\bibfnamefont{H.~B.} \bibnamefont{Su}},
  \bibinfo{author}{\bibfnamefont{J.~W.} \bibnamefont{Davenport}},
  \bibnamefont{and} \bibinfo{author}{\bibfnamefont{J.}~\bibnamefont{Tafto}},
  \bibinfo{journal}{Physical Review B} \textbf{\bibinfo{volume}{69}},
  \bibinfo{pages}{064501} (\bibinfo{year}{2004}).

\end{thebibliography}

\newpage
\begin{table}[h!]
\begin{tabular}[c]{|c|c|c|c|c|c|}
\hline
Sample & $T_c\ [K]$ & $T_s\ [^\circ C]$ & $t_s\ [h]$ & a [\AA] & c [\AA] \\
\hline
\MB\ & 35.0 & 1200 & 96 & 3.082 & 3.521 \\
\hline
\MBCa\ & 32.0 & 1200 & 48 & 3.069 &  3.520\\
\hline
\MBCb\ & 24.5 & 950 & 1 & 3.047 & 3.519 \\
\hline
\MAbB\ & 29.3 & 1200 & 96 & 3.077 & 3.476 \\
\hline
\MAcB\ & 7.2 & 1200 & 96 & 3.066 & 3.424 \\
\hline
\end{tabular}

\caption{Transition temperature ($T_c$), Sintering temperature ($T_{s}$), Sintering time ($t_s$), and lattice parameters (a,c) for the different samples. }\label{Tab:Samples}
\end{table}

\newpage

\newpage

\newpage

\newpage

\newpage

\newpage

\newpage

\newpage

\newpage

\newpage

\end{document}